\documentclass{aa}  
\bibpunct{(}{)}{;}{a}{}{,}
\usepackage{graphicx}
\usepackage{txfonts}
\usepackage{amsmath}
\usepackage{amssymb} 
\usepackage{rotating}
\usepackage{multirow}
\begin{document}

   \title{Revealing a hard X-ray spectral component that reverberates within one light hour of the central supermassive black hole in Ark 564}


   \author{M. Giustini
          \inst{1,2,3}
          \and
          T. J. Turner\inst{4}
          \and
          J. N. Reeves\inst{5,4}
          \and
          L. Miller\inst{6}
          \and
          E. Legg\inst{6}
          \and
          S. B. Kraemer\inst{7}
          \and
          I. M. George\inst{4}
          }

   \institute{SRON Netherlands Institute for Space Research, Sorbonnelaan 2, 3584 CA Utrecht, the Netherlands\\
              \email{m.giustini@sron.nl}
         \and
             XMM-Newton Science Operations Centre, ESA/ESAC, Villafranca del Castillo, Apartado 78, 28692 Villanueva de la Ca\~nada, Spain\\
         \and
                Center for Space Science and Technology, University of Maryland, Baltimore County, 1000 Hilltop Circle, Baltimore, MD 21250, USA\\
        \and
                Department of Physics, University of Maryland Baltimore County, Baltimore, MD 21250, USA\\
        \and
                Astrophysics Group, School of Physical and Geographical Sciences, Keele University, Keele, Staffordshire ST5 5BG, UK\\
        \and
                Department of Physics, University of Oxford, Denys Wilkinson Building, Keble Road, Oxford, OX1 3RH, UK\\
        \and
                Institute for Astrophysics and Computational Sciences, Department of Physics, The Catholic University of America, Washington, DC 20064, USA
             }

   \date{ }
\titlerunning{X-ray reverberation within one light hour of the central SMBH in Ark 564}
\authorrunning{M. Giustini et al.}

 
  \abstract
   {Arakelian 564 (Ark 564, $z=0.0247$) is an X-ray-bright narrow-line Seyfert 1 galaxy. By using advanced X-ray timing techniques,
an excess of ``delayed'' emission in the hard X-ray band ($4-7.5$ keV) following about 1000 seconds after ``flaring'' light in the soft X-ray band ($0.4-1$ keV) was recently detected. }
   {We report on the X-ray spectral analysis of eight XMM-Newton and one Suzaku observation of Ark 564. Our aim is to characterise the X-ray spectral properties of the source in the light of these
recently reported results.  }
   {High-resolution spectroscopy was performed with the RGS in the soft X-ray band, while broad-band spectroscopy was performed with the EPIC-pn and XIS/PIN instruments. We analysed time-averaged, flux-selected, and time-resolved spectra.}
   {Despite the strong variability in flux during our observational campaign, the broad-band spectral shape of Ark 564 does not vary
dramatically 
and can be reproduced either by a superposition of a power law and a blackbody emission or by a Comptonized power-law emission model. High-resolution spectroscopy revealed ionised gas along the line of sight at the systemic redshift of the source, with a low column density ($N_H \sim 10^{21}$ cm$^{-2}$) and a range of ionisation states ($-0.8 < \log(\xi/$erg cm s$^{-1}) < 2.4$). Broad-band spectroscopy revealed a very steep intrinsic continuum (photon index $\Gamma\sim 2.6$) and a rather weak emission feature in the iron K band (EW $\sim 150$ eV); modelling this feature with a reflection component requires highly ionised gas, $\log(\xi/$erg cm s$^{-1})> 3.5$.
A reflection-dominated or an absorption-dominated model are similarly able to well reproduce the time-averaged data from a statistical point of view, in both cases requiring contrived geometries and/or unlikely physical parameters.
Finally, through time-resolved analysis we spectroscopically identified the ``delayed'' emission as a spectral hardening above $\sim 4$ keV; the most likely interpretation for this component is a reprocessing of the ``flaring'' light by gas located at 10-100 $r_g$ from the central supermassive black hole that is so hot that it can Compton-upscatter the flaring intrinsic continuum emission.}
   {}


   \maketitle
%

\section{Introduction}
Arakelian 564 (Ark 564) is one of the X-ray-brightest AGN of the sky, with a 2-10 keV flux $f_{2-10}\sim 10^{-11}$ erg s$^{-1}$ cm$^{-2}$.
It is classified as narrow-line Seyfert 1 (NLS1), with a FWHM(H$\beta$)$\sim$750 km s$^{-1}$ \citep{1990A&AS...85.1049S}.
Like many NLS1s, Ark 564 displays an extremely variable X-ray flux and a very steep spectral shape \citep{1994MNRAS.271..958B, 2001ApJ...561..131T,2002ApJ...568..610E}. Possible complexities in its X-ray spectrum related to the reprocessing of the primary continuum emission onto ionised matter have been reported by \citet{1999MNRAS.308L..34V}, \citet{1999ApJ...526...52T}, \citet{2001A&A...365..400C}, \citet{2004MNRAS.347..854V}, \citet{2004ApJ...603..456M}, \citet{2007A&A...461..931P}, \citet{2007ApJ...671.1284D}, \citet{2008A&A...490..103S}.

\citet{2012ApJ...760...73L} exploited a long XMM-Newton observational campaign to perform a detailed timing analysis of the X-ray properties of Ark 564. The main result was the discovery of an excess of emission in the hard (4-7.5 keV) band following soft (0.4-1 keV) flares that appeared $\sim 1000$ s after the flares and lasted $\sim 1500$ s. This is consistent with light reverberation on some scattering material located $\sim 100 r_g$ from the continuum source ($r_g\equiv GM_{BH}/c^2$).

The main goal of this work is characterising the X-ray spectral properties of Ark 564 in the light of the results reported by \citet{2012ApJ...760...73L}. The paper is organised as follows: the observations are presented in Sect. \ref{obs}, the spectral analysis is reported in Sect. \ref{analysis}, the results are summarised in Sect. \ref{results}, and conclusions are drawn in Sect. \ref{concl}.

\section{Observations and data reduction}\label{obs}
In the following subsections we report on the data reduction of eight XMM-Newton and one Suzaku X-ray observations of Ark 564.
\subsection{ XMM-Newton observations}

\begin{table*}
\caption{Summary of XMM-Newton observations for Ark 564\label{table1}} 
\centering
\begin{tabular}{c c c c c c}
\hline\hline
OBSID  & Start date  & Duration & Instrument & Net exposure & Excision \\
\hline
\multirow{3}{*}{0670130201} & \multirow{3}{*}{05/24/2011} & \multirow{3}{*}{59.5}     & pn   & 38.6    & 8   \\
 & & & RGS1 & 55.6 &\multirow{2}{*}{/}\\
 & & & RGS2 & 55.5 & \\
  & & & & & \\
\multirow{3}{*}{0670130301} & \multirow{3}{*}{05/30/2011} & \multirow{3}{*}{55.9}     & pn   & 34.3  & 0     \\
 & & & RGS1 & 50.2&\multirow{2}{*}{/}\\
 & & & RGS2 & 50.1 & \\ 
  & & & & & \\
\multirow{3}{*}{0670130401} & \multirow{3}{*}{06/05/2011 } & \multirow{3}{*}{63.6}     & pn   & 31.1    & 0   \\
 & & & RGS1 & 43.0 &\multirow{2}{*}{/}\\
 & & & RGS2 & 42.9 &\\ 
  & & & & & \\
   \multirow{3}{*}{0670130501} & \multirow{3}{*}{06/11/2011} & \multirow{3}{*}{67.3}     & pn   & 40.6   & 5   \\
 & & & RGS1 & 59.7&\multirow{2}{*}{/}\\
 & & & RGS2 & 59.6 & \\
  & & & & & \\
 \multirow{3}{*}{0670130601} & \multirow{3}{*}{ 06/17/2011} & \multirow{3}{*}{60.9}     & pn   & 33.5   &  0  \\
 & & & RGS1 & 48.4 &\multirow{2}{*}{/}\\
 & & & RGS2 & 48.2 & \\
  & & & & & \\
 \multirow{3}{*}{0670130701} & \multirow{3}{*}{06/25/2011 } & \multirow{3}{*}{64.4}     & pn   & 29.1    & 0   \\
 & & & RGS1 & 42.1&\multirow{2}{*}{/}\\
 & & & RGS2 & 41.9&\\
  & & & & & \\
 \multirow{3}{*}{0670130801} & \multirow{3}{*}{06/29/2011} & \multirow{3}{*}{58.2}     & pn   & 40.5   & 4    \\
 & & & RGS1 & 58.1&\multirow{2}{*}{/}\\
 & & & RGS2 & 58.0 &\\
  & & & & & \\
 \multirow{3}{*}{0670130901} & \multirow{3}{*}{07/01/2011} & \multirow{3}{*}{55.9}     & pn   & 38.7     & 5  \\
 & & & RGS1 & 55.7&\multirow{2}{*}{/}\\
 & & & RGS2 & 55.7&\\    
\hline
\end{tabular}
\tablefoot{Observation ID, start date, duration of the observation (in kiloseconds), net exposure time after background flaring removal (in kiloseconds; the pn values include the loss of exposure due to the small window observing mode live time), and excision radius used to correct for pile-up (in arcseconds).}
\end{table*}

Ark 564 was observed by XMM-Newton eight times between May and July 2011. 
Data were processed using \textsc{HEAsoft v.6.12} and \textsc{SAS v.11.0.0}.

The reflection grating spectrometer (RGS) data were processed using calibration files generated in June 2012.
Background light curves were generated and inspected for the presence of flares: a uniform threshold of  $0.1$~ct~s$^{-1}$ for the background rate was then used to generate good time interval tables for each observation. Net exposure times after background flaring removal are reported in Table~\ref{table1}.
First-order spectra and response matrices were extracted using the \textit{rgsproc} task using source coordinates as given by NED (RA=340.663939, dec=+29.725364) as input to prevent possible offsets between different observations.
While there is considerable flux variability between the different observations, the spectral shape is found not to vary, therefore we proceeded with stacking the eight observations using the task \textit{rgscombine}. We obtained two high signal-to-noise spectra with $\sim 412$ ks of exposure each, and a total of $\sim 10^6$ counts for the combination of the two instruments, RGS1 and RGS2. Source spectra were rebinned by a factor of 3. 

All the EPIC pn observations were performed in small window mode and used the thin optical blocking filter. 
The raw ODF pn event files were filtered, and we retained only best-quality (FLAG=0) single and double (PATTERN $\leq 4$) events. 
Source events were extracted from circular regions with a 36$''$ radius, while background events were extracted from two 
boxes of $61'' \times 100''$ and $51'' \times 51''$  in size. 
Background light curves were then extracted for single events alone in the $10-12$~keV band for each observation. 
After inspecting these light curves,  we removed strong flaring background time intervals using a conservative threshold of $0.1$~ct~s$^{-1}$. The net exposure times after the flaring background removal is reported in Table~\ref{table1}. These values also account for the small window mode live time of $\sim 71 \%$.

The \textsc{SAS} task \textit{epatplot} was then used to estimate the fraction of pile-up affecting the data.
By using the standard energy interval $0.5-2$~keV to estimate the pile-up fraction, excellent numerical results were obtained, with the single or double pattern observed distributions deviating by less than 1\% from the predicted ones. However, a visual inspection of the pattern distribution revealed significant deviations both at energies below 1 keV (with an observed excess of single events) and at energies above 1 keV (with an observed excess of double events), the two effects compensating in the numerical estimate for the pile-up. We then re-estimated the pile-up fraction by
focusing on the $1-9$~keV band. Using this band as a reference, we found non-negligible deviations (of the order of $3-5\%$) of the observed pattern distribution from the predicted one in half of the observations. The cores of the PSF were then excised using a radius proportional to the amount of pile-up in the data, until we were able to recover the predicted pattern distribution within a $1-2\%$ uncertainty over the whole energy range used in this work, that is, $0.4-10$~keV. Excision radii are reported in the last column of Table~\ref{table1}. Redistribution matrices (RMF) and ancillary response files (ARF) were generated for each observation at the position of the source using tasks \textit{rmfgen} and \textit{arfgen} , taking into account the excision of the PSF core.

The EPIC-MOS data were heavily affected by pile-up and so were not useable.

\subsection{Suzaku observation}
Suzaku \citep{2007PASJ...59S...1M} carries four co-aligned telescopes  that focus X-rays onto a suite of CCD cameras comprising the X-ray Imaging Spectrometer \citep[XIS][]{2007PASJ...59S..23K}.
XIS CCDs 0, 2, 3 are front-illuminated (FI) and  XIS\,1 is  back-illuminated. 
XIS\,1 has an enhanced soft-band response relative to the FI units, but has a  lower effective area at 6\,keV and  a higher background level at high energies.  XIS\,2 failed on 2006 November 9 and hence was not used in our analysis.
Suzaku also carries a non-imaging, collimated  Hard X-ray Detector \citep[HXD,][]{2007PASJ...59S..35T}, whose PIN detector provides useful AGN data typically over the range 15-70\,keV.

A Suzaku  observation of Ark~564 was made in 2007 June 26$-$28  (OBSID 702117010), and the data  were  reduced using v6.16 of {\sc HEAsoft}. 
We screened the XIS and PIN events to exclude data taken during passage through the South Atlantic Anomaly, starting and ending within 500 seconds of entry or exit.  In addition, we excluded data with an Earth  elevation angle smaller than 5$^\circ$. 
Ark~564 was observed at the nominal centre position for the HXD. 
FI CCDs used  $3 \times 3$ and $5 \times 5$ edit-modes, with normal clocking mode. 
Good events (with grades 0, 2, 3, 4, and 6) were selected and  hot or flickering pixels removed using the {\sc SISCLEAN} script.  
The spaced-row charge  injection (SCI) was used. 
The net exposure times were 70 ks per XIS unit and 81 ks for the PIN.
XIS products were extracted from circular regions of 2.9\arcmin \, radius while background spectra were extracted from a region
of the same size, offset from the source and avoiding the calibration sources in the chip corners.

For PIN analysis we used the model ``D'' background (\footnote[1]{http://www.astro.isas.jaxa.jp/suzaku/doc/suzakumemo/ suzakumemo-2007-01.pdf}). We checked the correctness of the model adopted by comparing it to the Earth-occulted data, finding an overestimation of the NXB of the level of 11.5\%; this factor was taken into account into the subsequent analysis.
Following standard practice, the time filter resulting from screening the observational data was applied to the background events model.  
The ftool {\sc hxdpinxbpi} was used to create a PIN background spectrum from the screened background data, taking the shape and flux level of the cosmic X-ray background 
\citep{1987PhR...146..215B,1999ApJ...520..124G} in the Suzaku PIN field of view into account.  
We used the response file  ae$\_$hxd$\_$pinhxnome3$\_$20080129.rsp for spectral fitting.

 During the  Suzaku observation Ark 564  was found to have source count rates
2.132 $\pm$ 0.006 (summed XIS 0, 3 over 0.75-10\,keV)  and 0.010 $\pm 0.002$ (PIN over 15-50\,keV) ct/s.
The background was $1\%$ of the total XIS count rate in the full band.
The source comprised 3\% of the total counts in the PIN band.

 Spectral fits used data  from XIS 0 and 3, in the energy range $0.75-10$\,keV
and from the PIN over $15-50$\,keV.
Data in the range 1.75-1.9\,keV were excluded from the XIS spectral analysis owing to
uncertainties in calibration around the instrumental Si K edge.
XIS\,1 was not used owing to the
higher background level at high energies.
The data were binned to have at least 20 counts per bin.
In the spectral fitting, the PIN flux was scaled by a
factor 1.18, which is appropriate for the cross calibration
of the XIS and PIN for this  observation epoch
\footnote[2]{ftp://legacy.gsfc.nasa.gov/suzaku/doc/xrt/suzakumemo-2008-06.pdf}. A systematic error of 3\% was added to the data.

\section{Spectral analysis}\label{analysis}
We used Xspec v.12.7.0 for the spectral analysis. We used the $\chi^2$ statistics as a goodness-of-fit measure, in a valid regime assured by the high number of counts in each energy channel for each analysed spectrum. All the models include a Galactic column density $N_H=5.43\times 10^{20}$ cm$^{-2}$ \citep{2005A&A...440..775K}, modelled with \texttt{tbabs} for the low-resolution pn data \citep{2000ApJ...542..914W}, with \texttt{tbnew} for the high-resolution RGS data (Wilms et al. 2015, in preparation). Errors are given at the 90\% confidence level ($\Delta\chi^2=2.706$ for one parameter of interest). A cosmological redshift $z=0.02468$ was adopted for the source. We first analysed the RGS data to constrain the physical properties of any possible soft-band absorbing or emitting component (Sect.~\ref{RGS}); the results obtained for the soft band were then used as input in our modelling of the full-band EPIC pn and Suzaku XIS/PIN spectra (Sect.~\ref{PN}).

\subsection{Grating analysis\label{RGS}}
\begin{table*}
\caption{Fit to the RGS continuum\label{table2}} 
\centering
\begin{tabular}{c c c c c c c c}
\hline\hline
Model &\multicolumn{6}{c}{Parameters} & Statistics \\
\hline
 \multirow{3}{*}{bbody + po} & $\Gamma$ & $N_{1\, keV}/10^{-2}$ & $kT$/eV & $N_{bb}/10^{-4}$ &   &  & $\chi^2/\nu$\\
 & $2.78^{+0.02}_{-0.02}$  & $1.78^{+0.02}_{-0.02}$  & $128^{+2}_{-2}$  & $3.2^{+0.2}_{-0.2}$ & & & 2596/1295 \\
         \multicolumn{8}{c}{}\\
\multirow{2}{*}{optxagnf} & $\Gamma$ & $f_{pl}$ & $L/L_{Edd}$ & $kT$/eV & $\tau$ & $r_{cor}/r_g$ & $\chi^2/\nu$\\
         & $2.48^{+0.13}_{-0.03}$ & $0.60^{+0.16}_{-0.06}$ & $1.12^{+0.07}_{-0.02}$ & $168^{+13}_{-12}$ & $24^{+4}_{-3}$ & $80^{+6}_{-24}$ & 2594/1293\\
\hline
\end{tabular}
\tablefoot{The power-law normalisation $N_{1\,keV}$ is listed
in units of photons keV$^{-1}$ cm$^{-2}$ s$^{-1}$ at 1 keV; the blackbody normalisation $N_{bb}=L_{39}/D_{10}^2$, where $L_{39}$ is the luminosity in units of $10^{39}$ erg s$^{-1}$ and $D_{10}$ is the source distance in units of 10 kpc.}
\end{table*}

The RGS1 and RGS2 stacked spectra were simultaneously fitted over the $0.4-1.9$~keV band. Bad channels were ignored, and a systematic error of 3\% was added  in quadrature with the photon shot noise to account for the calibration uncertainties between the two RGS \citep[see e.g.][]{2011A&A...534A..37K}. 

The spectral shape in this energy range is extremely steep, and a simple power-law emission model gives a very poor fit to the data ($\chi^2/\nu=4700/1297$). 
Adding a blackbody component significantly improves the fit statistic (Table \ref{table2}). The black hole mass of Ark 564 is quite low \citep[$M_{\rm{BH}}\sim 2.6\times 10^6 M_{\odot}$,][]{2004AJ....127.3168B}, so we might expect the tail of the thermal emission from the accretion disk to emerge in the soft X-ray band - but a simple blackbody emission component is obviously only a phenomenological parametrization for the continuum shape.

We replaced the \texttt{power law + blackbody} model with \texttt{optxagnf}, which represents the expected emission spectrum from the inner accretion flow around a black hole \citep{2012MNRAS.420.1848D}. This model assumes that the accretion luminosity (parametrized in terms of the mass accretion rate $\dot{M}_{acc}$ and the black hole spin $a$) is produced in a disk + corona system. A geometrically thin and optically thick accretion disk extends from an outer radius $r_{out}$ up to the innermost stable circular orbit around the black hole and emits as a colour-temperature corrected blackbody (i.e. it has an effective temperature higher than a standard SS disk at small radii within the region where electron scattering starts to dominate the disk opacity: this effect is strongest for the hottest accretion disks, such as for low black hole mass and high mass accretion rate AGN) up to a radius $r_{cor}$.
Inside this radius, the disk seed photons are Compton-upscattered in a two-phase hot medium: a fraction of them $f_{pl}$ is Comptonized on a very hot and optically thin medium (e.g. the ``X-ray corona''), while the remaining $(1 - f_{pl} )$ is Comptonized on a much colder and optically thick medium (e.g. the ``inner disk'' or a ``disk atmosphere'' or a ``disk skin'').  We fixed the black hole mass $M_{BH}=2.6\times 10^6 M_{\odot}$ \citep{2004AJ....127.3168B}, the black hole spin $a=0$\footnote{The spin parameter $a$ is degenerate with the mass accretion rate unless one can constrain the latter with UV flux measurements from the disk \citep[see e.g.][]{2013MNRAS.434.1955D}, which we lack because of strong contamination of the host galaxy emission in this band; therefore we fixed for simplicity $a=0$.}, and the accretion disk outer radius $r_{out}=10^5 r_g$. The free parameters are then the mass accretion rate $\dot{M}_{acc}$, the outer radius of the Compton-upscattering region $r_{cor}$, the temperature $kT$ and optical depth $\tau$ of its cold and optically thick portion, the fraction $f_{pl}$ of the energy that is upscattered in its hot and optically thin portion, and the observed photon index $\Gamma$.  
The \texttt{optxagn} model is also able to improve the fit statistics with respect to the power-law component alone and gives a statistically equivalent fit to the data as the \texttt{power law + blackbody} model (Table \ref{table2}); however, given the two more free parameters required by  \texttt{optxagn} and given the much faster performance of \texttt{power law + blackbody}, we proceeded to search for spectral residuals using the latter as the baseline model for the continuum.
First, the residuals were modelled using Gaussian emission or
absorption lines (Sect.~\ref{gauss}), then a fit using \texttt{Xstar} tables was performed (Sect.~\ref{xstar}).

 \subsubsection{Fit with Gaussians\label{gauss}}

   \begin{figure}
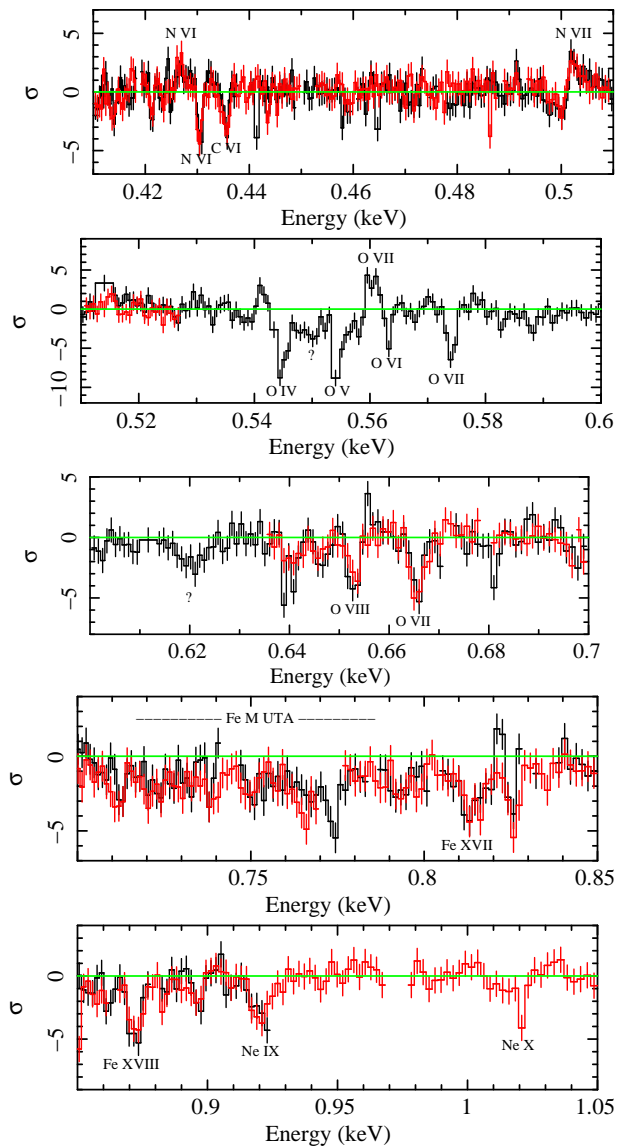

   \centering
    \includegraphics[ width=3cm,angle=-90]{figure1a.eps}
    \includegraphics[ width=3cm,angle=-90]{figure1b.eps}
        \includegraphics[ width=3cm,angle=-90]{figure1c.eps}
            \includegraphics[ width=3cm,angle=-90]{figure1d.eps}
                \includegraphics[ width=3cm,angle=-90]{figure1e.eps}
                          \caption{\label{Fs} RGS spectral residuals in the source rest frame. 
              }
   \end{figure}

A threshold of 99.9\% in statistical significance was adopted to consider the addition of a Gaussian line to the model as required by the data, that is, $\Delta\chi^2 > 16.3$ for three degrees of freedom associated with each line: centroid energy, width, and intensity. 
The analysis was simultaneously performed on the RGS1 and RGS2 dataset, and only features significant in both detectors were included in the model (except for energy ranges with missing chips, i.e. $E=0.90-1.18$ keV for the RGS1, and $E=0.52-0.62$ keV for the RGS2).

We detected 16 absorption and 3 emission lines in the soft X-ray spectrum of Ark 564: they are shown as residuals against the continuum in Fig.~\ref{Fs}, while their properties are listed in Table~\ref{Tgauss}.
For 14 out of 16 absorption lines we were able to associate a theoretical identification with H-like and He-like transitions of ions of C, N, O, Ne, and L-shell transitions of Fe: all of them are compatible with a zero-velocity shift. All the identified absorption lines are narrower than the instrumental RGS resolution except for the features observed between $\sim 700-800$ eV, which correspond to the Fe~I-XVI M-shell UTA \citep{2001ApJ...563..497B}, and the two high-ionisation transitions identified with Fe XVII and Ne IX, which display a FWHM $\sim 1000-1500$ km s$^{-1}$. The three emission lines are identified with H-like transitions of N~VI, N~VII, and O~VII. 

Two features remain unidentified at rest-frame energies of $\sim 552$ eV and $\sim 619$ eV; both of them are resolved by the RGS, but do fall in regions that are devoid of known strong spectral transitions of cosmically abundant elements. We note that within the measurements errors, they might be identified with N~VII and O~VII K$\alpha$ absorption by allowing a net blueshift of $\sim 0.1c$.

\begin{table*}
\caption{Absorption and emission lines detected in the RGS stacked spectra of ARK 564\label{Tgauss}}
\centering
\begin{tabular}{c c c c c c c}
\hline\hline
 $E_{\rm{rest}}$/eV &  $\sigma$/eV & FWHM/km s$^{-1}$ &  EW/eV & $\Delta \chi^2$ & ID  &  $E_{\rm{th}}$/eV\\
\hline
$427.3^{+0.4}_{-0.8}$ & $0.9^{+0.2}_{-0.2}$& $1490^{+330}_{-330}$ &  $0.34^{+0.11}_{-0.02}$ & 30.6 & N~VI & 430 (r), 426 (i), 420 (f)\\
$430.5^{+0.1}_{-0.3}$ & $<0.2$ & $<330$ & $-0.32^{-0.14}_{+0.03}$ & 83.3 & N~VI $1s-2p$& 431\\
$435.7^{+0.8}_{-0.5}$ & $<0.5$ & $<810$ & $-0.35^{-0.03}_{+0.17}$ & 41.4 & C~VI $1s-3p$  & 436\\
$502.6^{+0.7}_{-0.3}$ & $<1.2$& $< 1690$ & $0.31^{+0.08}_{-0.02}$ & 25.5 & N~VII $1s-2p$ & 500\\
$544.8^{+0.2}_{-0.5}$ & $<0.6$ & $<780$ & $-0.89^{-0.11}_{+0.12}$ & 104.2 & O~IV $1s-2p$ & 545 \\
$552.1^{+0.6}_{-2.2}$ & $3.8^{+1.2}_{-0.6}$ & $4860^{+1540}_{-770}$ & $-1.43^{-0.18}_{+0.46}$& 47.3 & ? &? \\
$554.4^{+0.3}_{-0.2}$ & $<0.4$ & $<510$ & $-0.77^{-0.10}_{+0.14}$& 45.8 & O V $1s-2p$ & 555 \\
$561.2^{+0.2}_{-0.7}$ & $<0.5$& $<630$  & $0.67^{+0.15}_{-0.13}$ & 55.1 & O~VII  & 574 (r), 569 (i), 561 (f)\\
$563.0^{+0.7}_{-0.3}$ & $<0.5$ & $<630$ &  $-0.33^{-0.17}_{+0.10}$ & 18.5 & O~VI $1s-2p$ & 564\\
$573.4^{+0.7}_{-0.3}$ & $<1.0$ & $<1230$ & $-0.48^{-0.22}_{+0.10}$ & 32.1 & O~VII $1s-2p (r)$ & 574\\
$619.4^{+2.4}_{-1.0}$ & $2.0^{+1.1}_{-1.3}$ & $2280^{+1260}_{-1480}$ & $-0.60^{-0.26}_{+0.26}$ & 17.8&  ? & ? \\
$652.4^{+0.9}_{-0.1}$ & $<0.5$ & $<530$ & $-0.33^{-0.26}_{+0.06}$ & 31.1&  O~VIII $1s-2p$ & 653\\
$665.9^{+0.2}_{-0.5}$ & $<0.5$ & $<530$ &  $-0.69^{-0.11}_{+0.12}$& 70.6 & O~VII $1s-3p$& 666\\
$745.8^{+10.6}_{-14.0}$ & $74.0^{+15.2}_{-12.0}$ & $70100^{+14400}_{-11370}$ & $-13.52^{-4.85}_{+3.20}$& 170.5 & Fe M UTA & $700-800$\\
$766.3^{+2.9}_{-6.8}$ & $2.4^{+4.6}_{-1.9}$ & $2210^{+4240}_{-1750}$ & $-0.72^{-0.51}_{+0.21}$& 20.2 & Fe M UTA & $700-800$\\
$812.9^{+1.7}_{-0.7}$ & $<2.2$ & $<1910$ & $-0.62^{-0.28}_{+0.21}$ & 17.4 & Fe~XVII $2p-3d$ & 812 \\
$871.2^{+1.5}_{-0.5}$ & $1.1^{+0.4}_{-0.5}$ & $890^{+330}_{-400}$ & $-1.29^{-0.15}_{+0.09}$ & 75.8 & Fe~XVIII $2p-3d$ & 873\\
$921.3^{+0.7}_{-1.6}$ & $2.1^{+1.1}_{-1.3}$& $1610^{+840}_{-1000}$ &  $-1.37^{-0.31}_{+0.32}$ & 53.8  & Ne~IX $1s-2p$& 922\\
$1019.7^{+2.0}_{-5.7}$ & $<4.4$& $<3050$ & $-1.03^{-0.46}_{+0.42}$ & 20.1  & Ne~X $1s-2p$& 1022\\
\hline
\end{tabular}
\tablefoot{Col.(1): rest frame line centroid energy; Col.(2): observed frame line width; Col.(3): observed frame line width; Col.(4):
observed frame line equivalent width; Col.(5): statistical improvement after adding the line, for three free parameters; Col.(6): line ID; Col.(7): theoretical line transition energy. All the errors are at the 90\% confidence level.}
\end{table*}


\begin{figure}[h!]
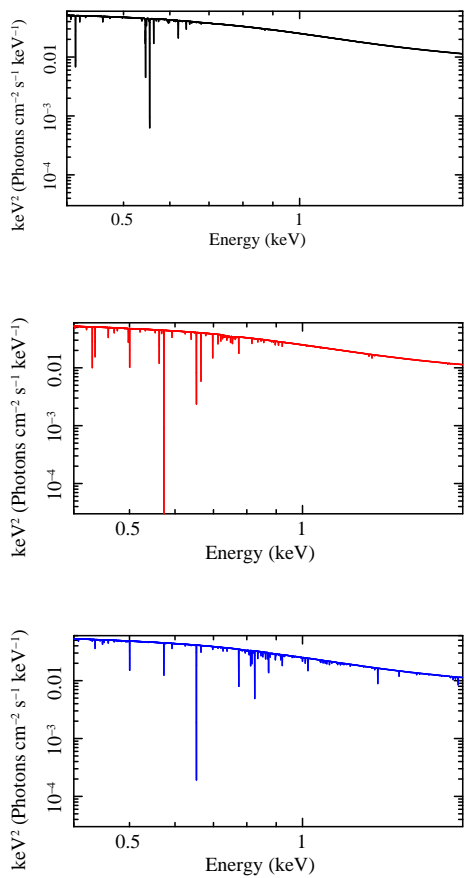

  \centering
    \includegraphics[angle=-90,width=0.35\textwidth]{figure2a.eps} 
    \includegraphics[angle=-90,width=0.35\textwidth]{figure2b.eps} 
    \includegraphics[angle=-90,width=0.35\textwidth]{figure2c.eps} 
    \caption{\label{Fxs} Rest-frame plot of the three zones of warm absorbing gas detected in the RGS data. From top to bottom, the ionisation parameter and column density are $\log(\xi/$erg cm s$^{-1})\sim -0.8,\,1.3,2.4$ and $\log (N_H/$cm$^{-2})\sim 20.1,\,20.3,\,20.9$.    }
\end{figure}
%

\subsubsection{Fit with \texttt{Xstar} tables\label{xstar}}

We generated a table using the latest version of \texttt{Xstar v.2.2.1bn3} and a finely spaced energy grid, using $10^5$ energy bins over the 
$0.1-20$~keV energy range used for the computation. The incident ionising continuum was a power law with a steep photon index $\Gamma=2.5$, appropriate for Ark 564. Solar abundances and a turbulent velocity of $100$~km~s$^{-1}$ were adopted. The gas column density was sampled over 10 points from $N_H=5 \times 10^{18}$~cm$^{-2}$ to $N_H=10^{21}$~cm$^{-2}$, while the gas ionisation state was sampled over 20 points from $\log(\xi/$erg cm s$^{-1})=-1.5$ up to $\log(\xi/$erg cm s$^{-1})=3.5$. 

The continuum emission was modelled with a power law plus a blackbody, and the model included the three emission lines detected in the previous section.
We obtained a fair fit to the data ($\chi^2_r=1.33$) using three zones of photo-ionised gas. We found no net velocity shift for any of the three zones. The relative contribution of each zone in terms of absorption components is shown in Fig.~\ref{Fxs}. The least ionised zone is responsible for most of the opacity for lowly ionised oxygen, that is, O~IV, O~V, and O~VI; the intermediate zone reproduces the H-like and He-like transitions of C~VI, N~VI-VII, and O~VII-VIII, as well as the least-ionised portion of the Fe M UTA; finally, the most-ionised zone contributes to filling the O~VIII H-like absorption as well as the higher ionisation states of the Fe M UTA, the Fe L, and the H-like and He-like Ne lines. Fit parameters are reported in Table~\ref{Txstar}.

We checked for partially covering gas along the line of sight by allowing any of the three absorbing zones to be only partially covering the source. While there is no statistical improvement when the zones are individually allowed to be partially covering, a $\Delta\chi^2=30$ for one extra degree of freedom is found when all the three zones share the same covering fraction $C_f=0.76^{+0.11}_{-0.02}$. While the gas ionisation state is not affected by introducing an unabsorbed component along the line of sight, the column densities measured in this scenario are of the order of 40\% higher than those measured in the scenario $C_f\equiv 1$.

\begin{table}
\caption{Fit to the RGS spectra with Xstar\label{Txstar}}
\centering
\begin{tabular}{cc}
\hline\hline
Parameter &  Value \\
\hline
$\Gamma$ & $2.81^{+0.02}_{-0.02}$\\
$N_{1\,keV}$ & $1.81^{+0.02}_{-0.02}$  \\
$kT$ & $136^{+2}_{-2}$ \\
$N_{bb}$ & $3.7^{+0.2}_{-0.2}$  \\
$\log N_{H,1}$ & $20.10^{+0.08}_{-0.09 }$ \\
$\log\xi_1$ & $-0.77^{+0.14}_{-0.15}$ \\
$\log N_{H,2}$ & $20.35^{+0.07}_{-0.07}$  \\
$\log\xi_2$ & $1.34^{+0.12}_{-0.08}$ \\
$\log N_{H,3}$ & $20.86^{+0.10}_{-0.08}$  \\
$\log\xi_3$ & $2.36^{+0.07}_{-0.06}$  \\
$\chi^2/\nu$ & 1770/1331 \\
\hline
\end{tabular}
\tablefoot{Units are keV$^{-1}$ cm$^{-2}$ s$^{-1}$ at 1 keV for the power law normalisation $N_{1\,keV}$; eV for the blackbody temperature $kT$; cm$^{-2}$ for the column densities $N_{H,i}$; erg cm s$^{-1}$ for the ionisation parameters $\xi_i$. The blackbody normalisation $N_{bb}=L_{39}/D_{10}^2$, where $L_{39}$ is the luminosity in units of $10^{39}$ erg s$^{-1}$ and $D_{10}$ is the source distance in units of 10 kpc.}
\end{table}

\subsection{Broad-band analysis\label{PN}}

\begin{figure*}[h!]
\begin{center}$
\begin{array}{cc}
    \includegraphics[angle=-90,width=0.4\textwidth, ]{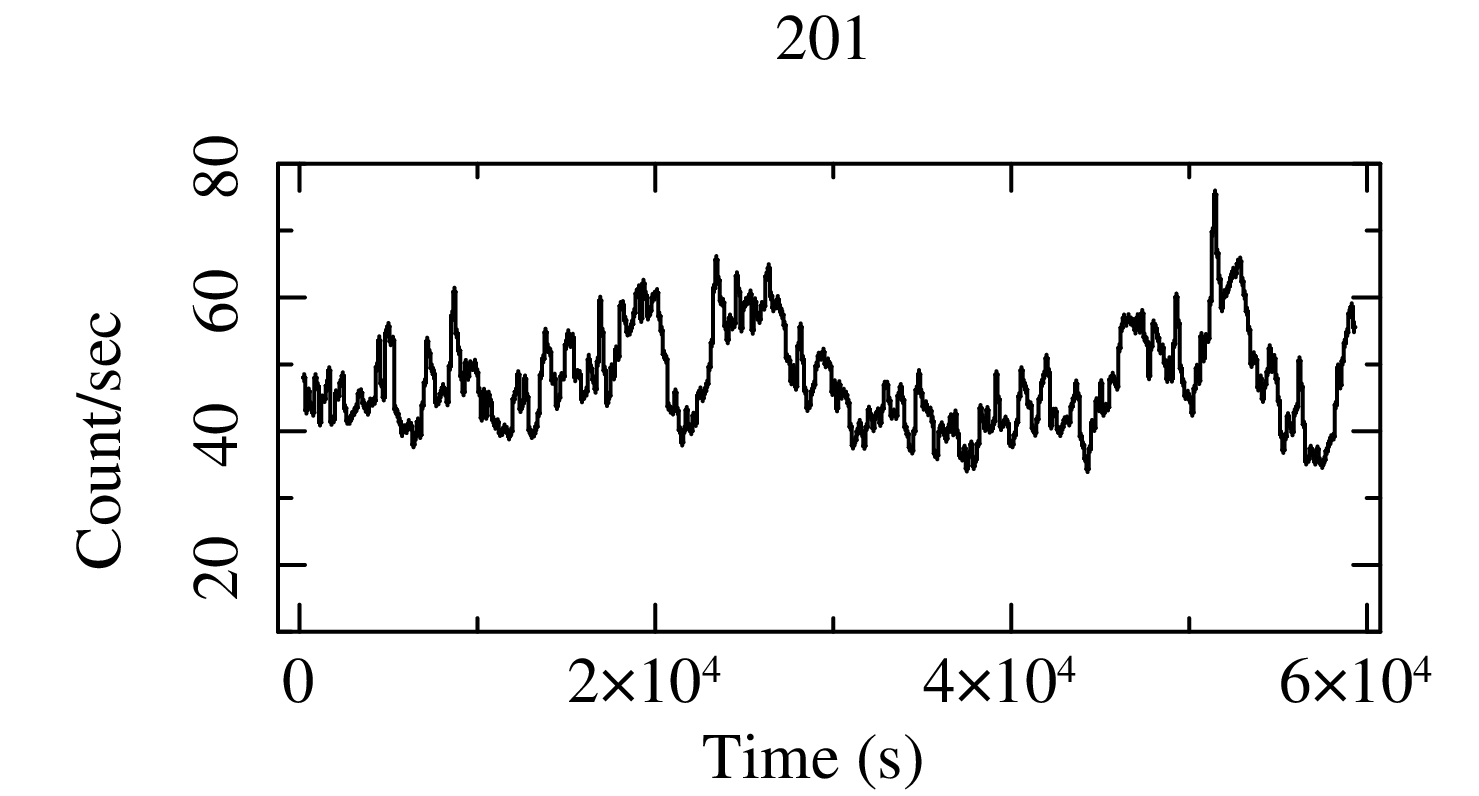} &
    \includegraphics[angle=-90,width=0.4\textwidth]{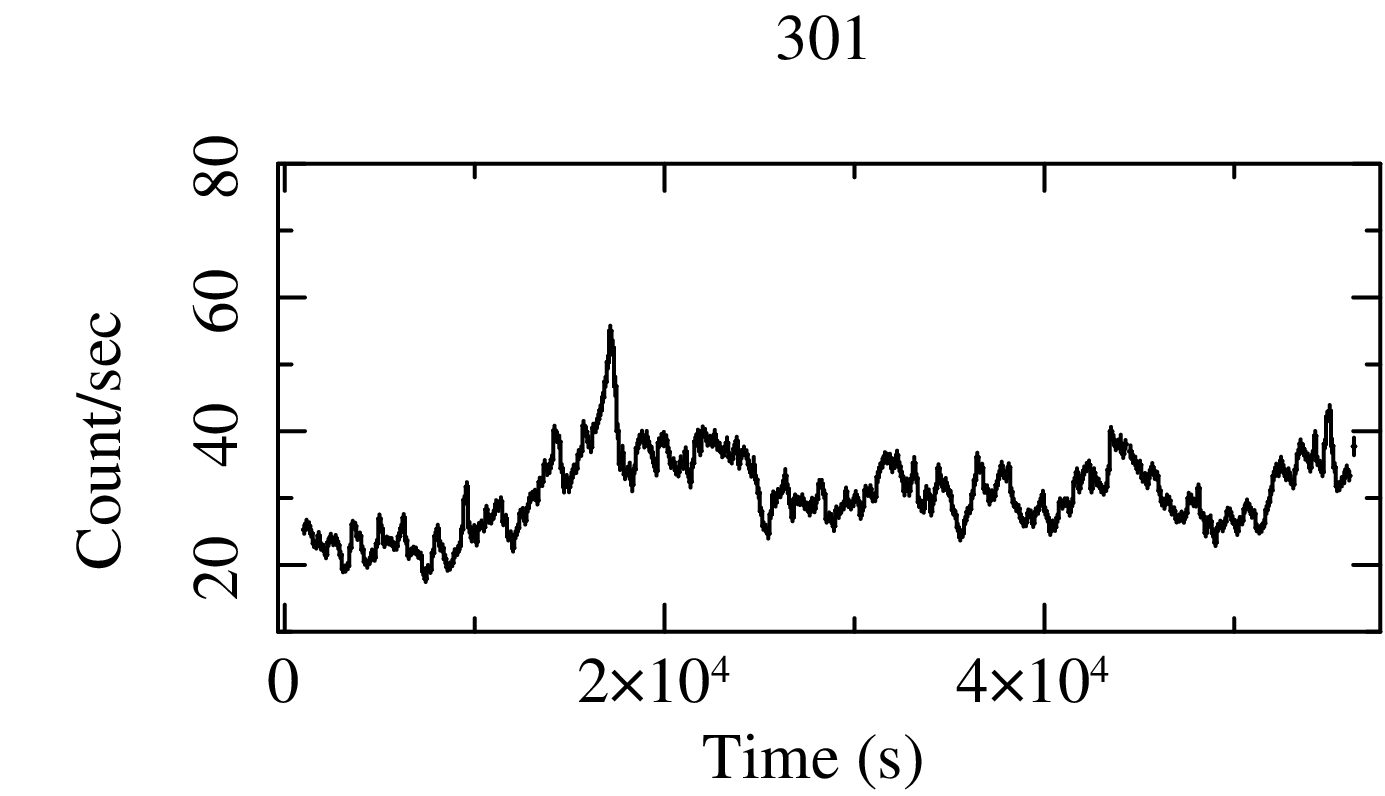} \\
    \includegraphics[angle=-90,width=0.4\textwidth]{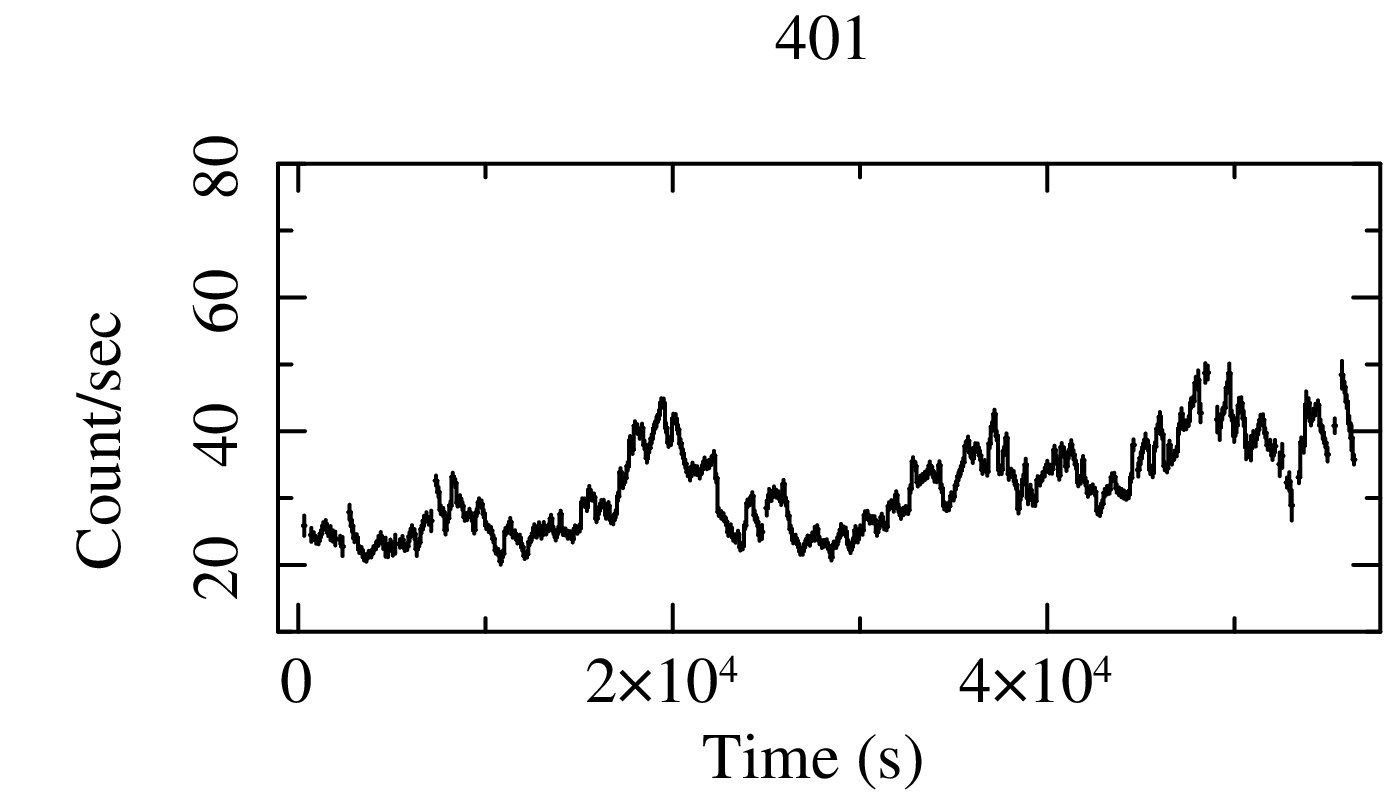} &
        \includegraphics[angle=-90,width=0.4\textwidth]{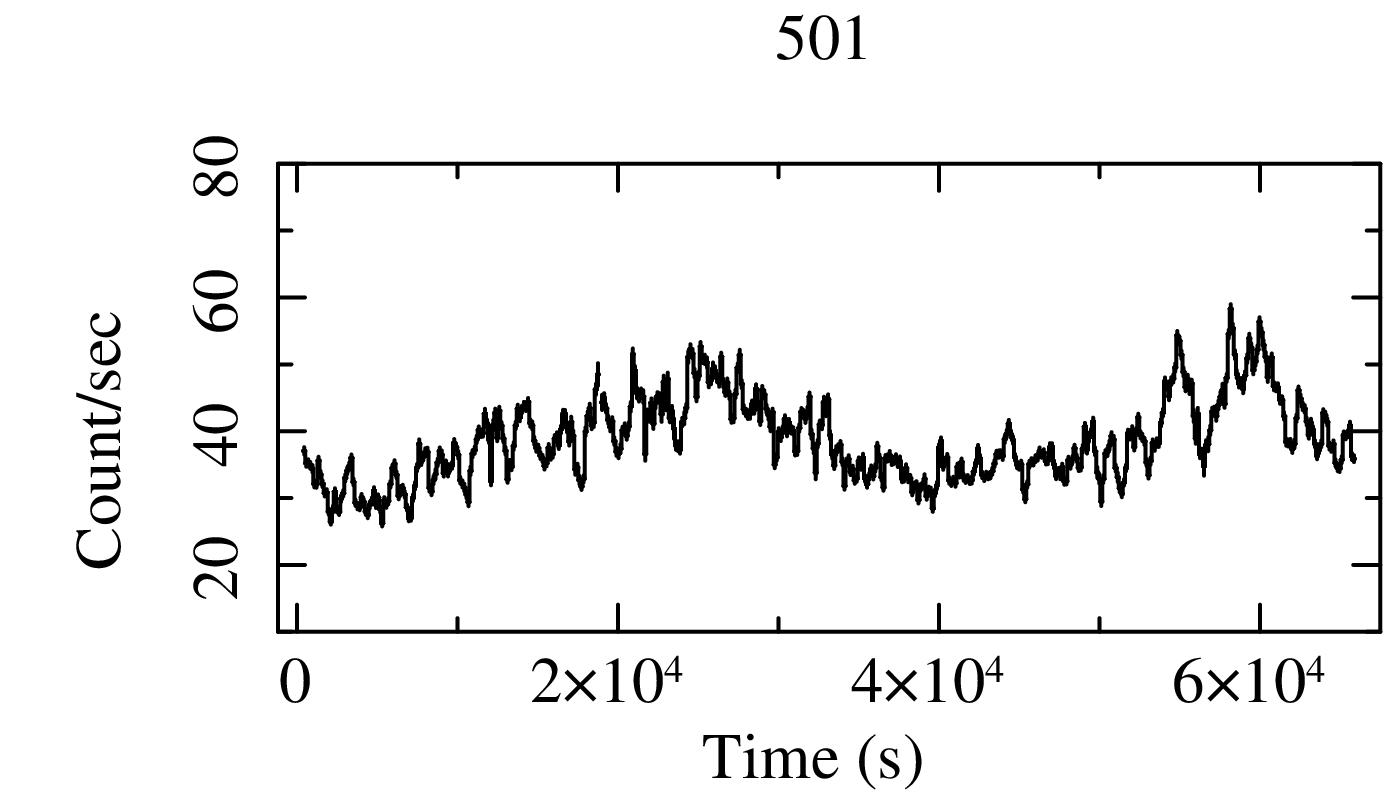} \\
    \includegraphics[angle=-90,width=0.4\textwidth]{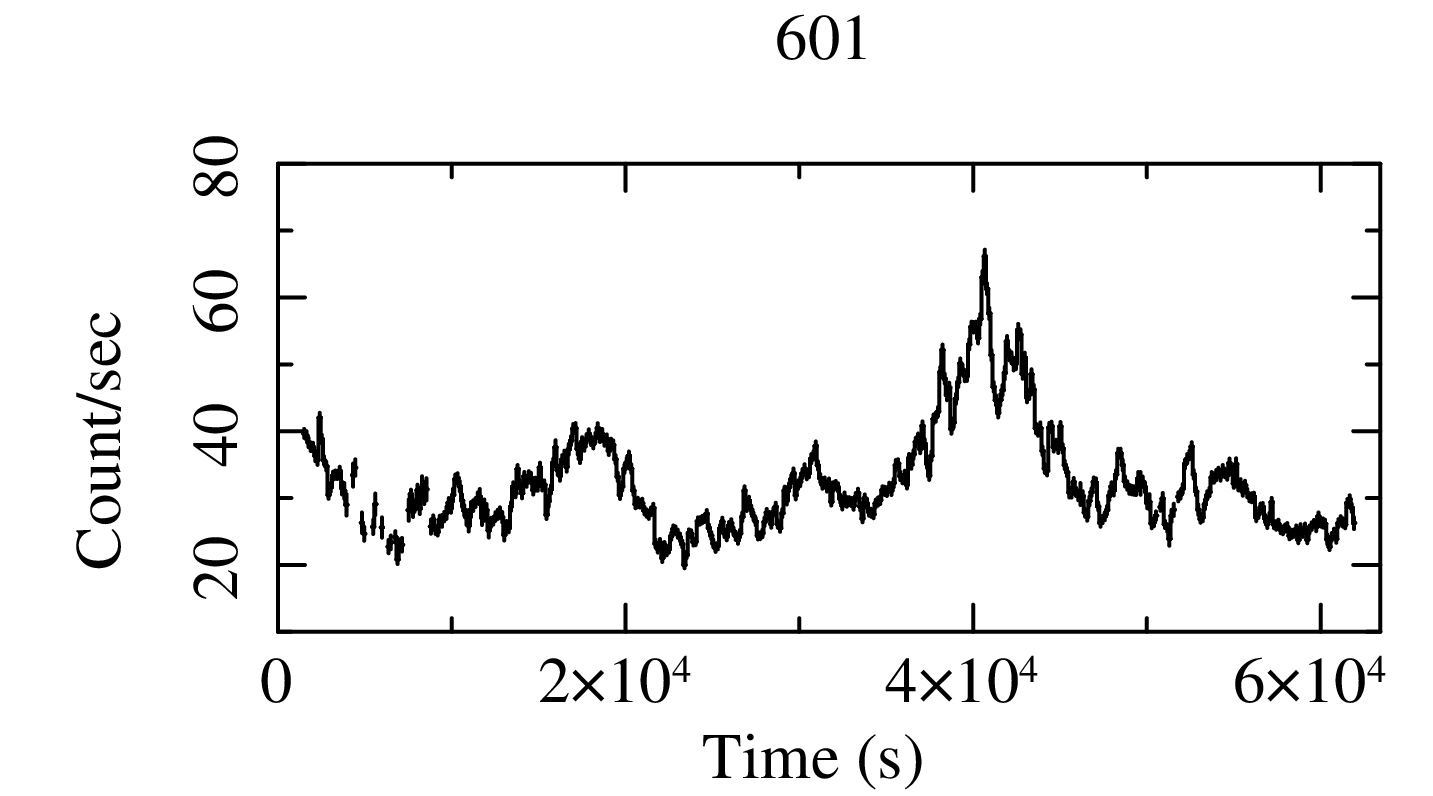} &
    \includegraphics[angle=-90,width=0.4\textwidth]{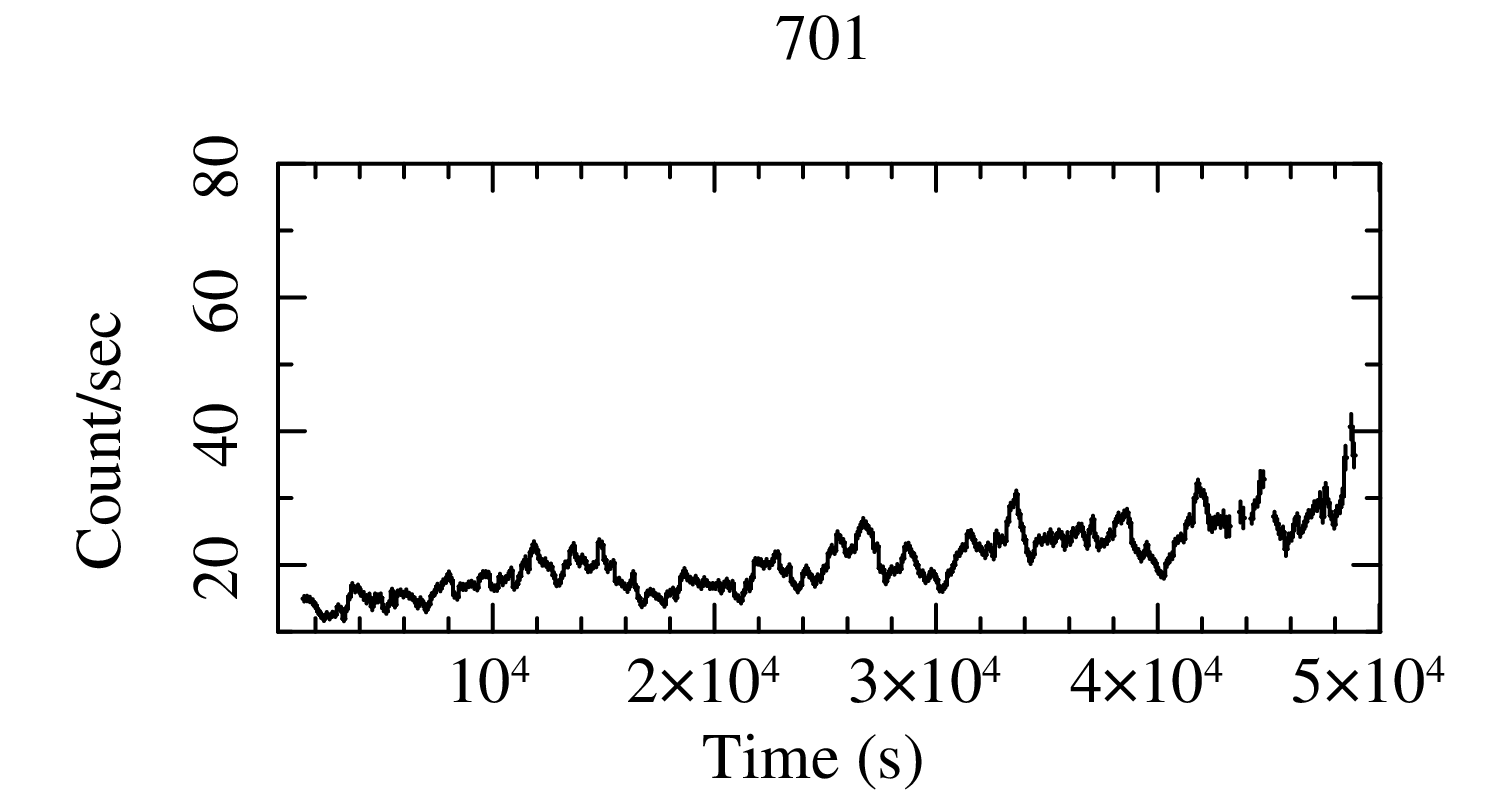} \\
    \includegraphics[angle=-90,width=0.4\textwidth]{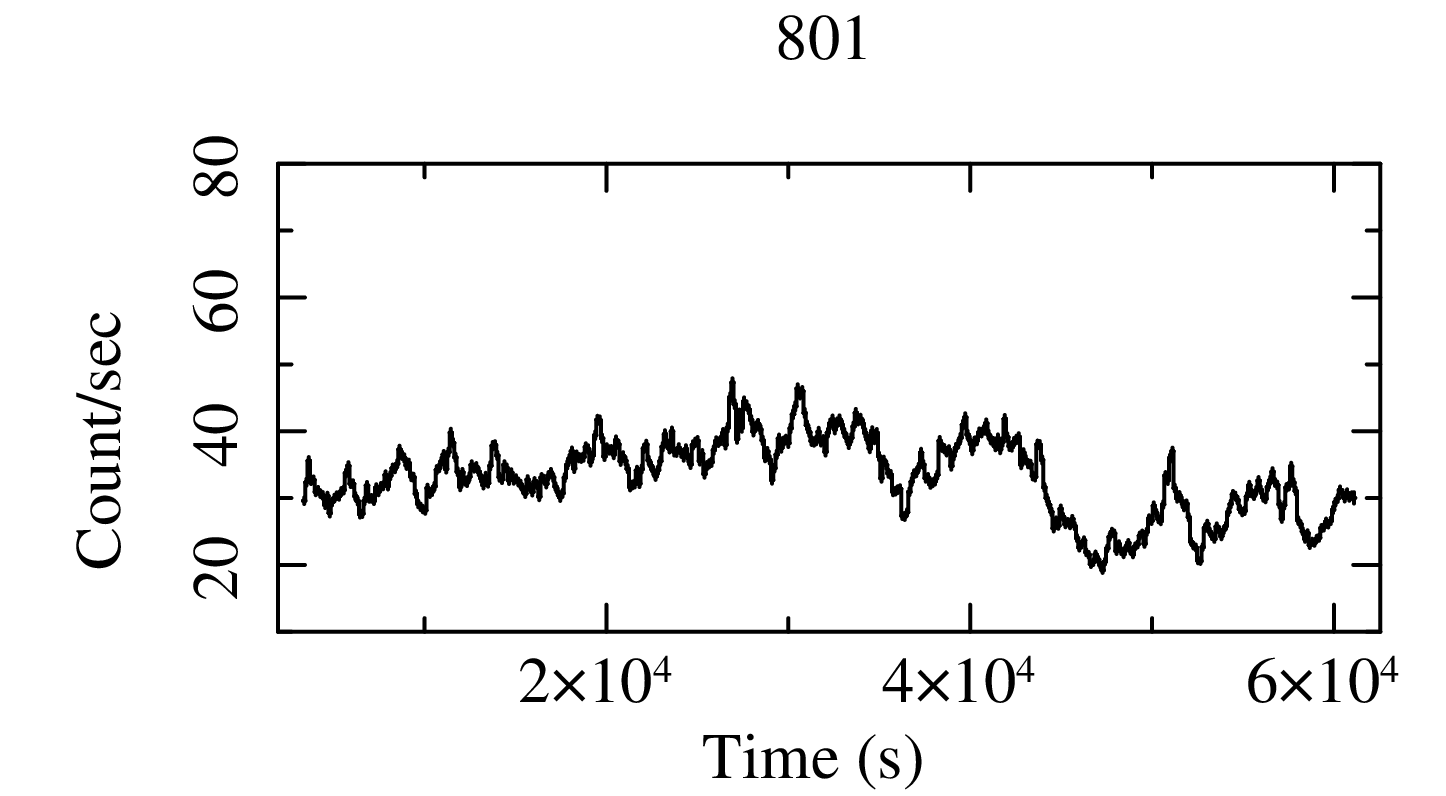} &
    \includegraphics[angle=-90,width=0.4\textwidth]{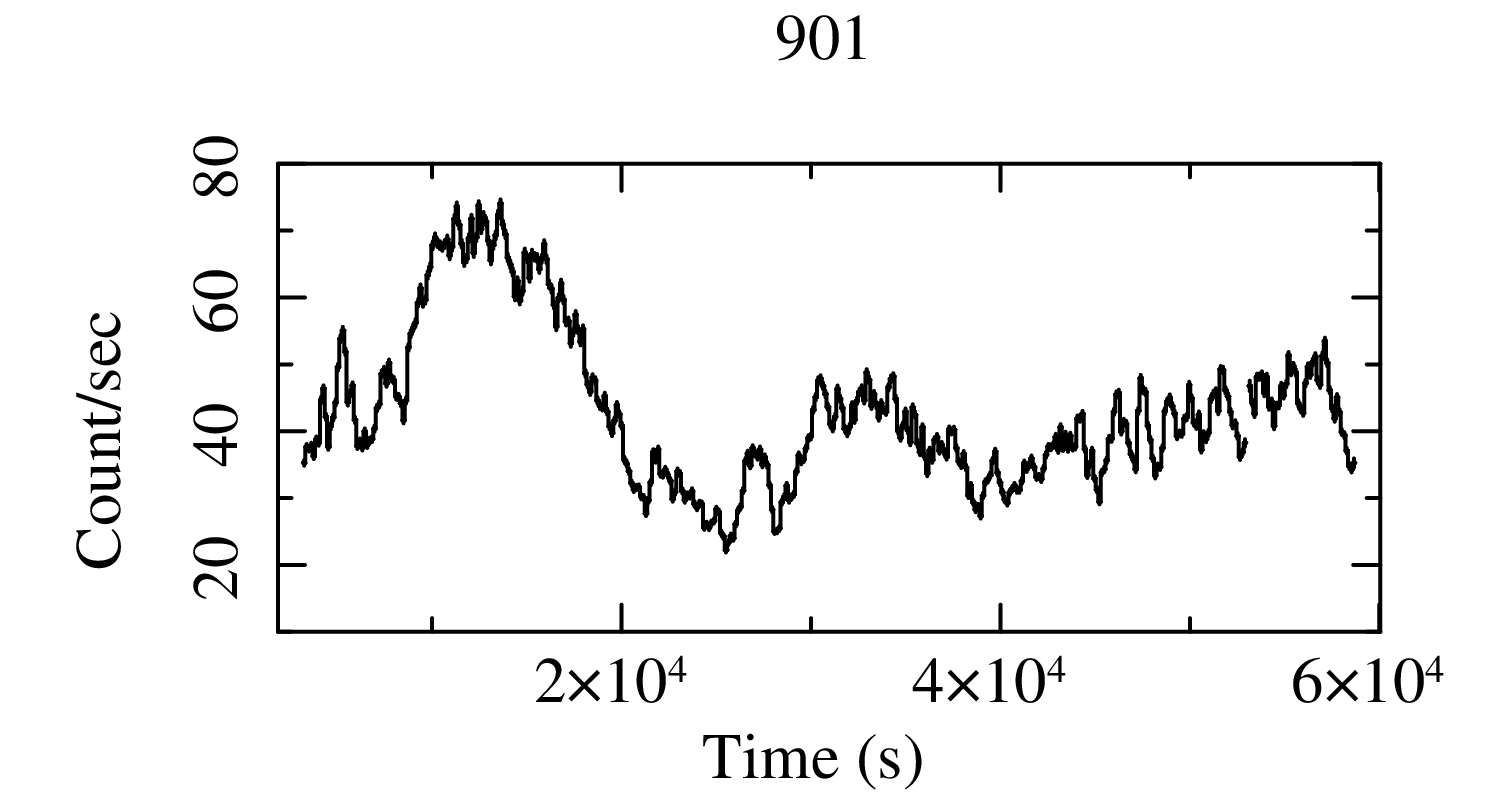} 
\end{array}$
\end{center}
    \caption{\label{Flc} Light curves for the eight XMM-Newton observations extracted in the $0.4-10$~keV band and binned to 128 s. 
    }
\end{figure*}

We show in Fig.~\ref{Flc} full-band (0.4-10 keV) pn light curves for each observation, binned to 128 s. The count rates for the four excised observations have been rescaled to account for the loss of counts due to discarding the central region of the PSF. Given the very low residual background count rate (typically lower than 1\% of the source one) after screening against the presence of strong flares, we show the source plus background light curves. The time variability of Ark 564 is striking on both short (kiloseconds) and long (hours and weeks) time scales. We first present the spectral analysis of the time-averaged spectra (Sect.~\ref{average}) and of the flux-selected spectra (Sect.~\ref{fluxsel}); we then present results of time-resolved spectral analysis (Sect.~\ref{timeres_FD}).

\subsubsection{Time-averaged spectra\label{average}}

The eight pn spectra were stacked using \textit{mathpha}. We generated an average background spectrum  by appropriately weighting each individual background spectrum by the BACKSCAL ratio between each source and background extraction region, which varies among the different observations because of the different excision radii used to clean the spectra from pile-up effects. Average RMF and ARF were created using \textit{addrmf} and \textit{addarf}.  Following \citet{2008A&A...483..437M}, each individual spectrum was binned at the HWHM instrumental energy resolution. A systematic error of 1\% was added to the data below 2 keV to account for uncertainties in the instrument calibration \footnote{There are typically $>>10^5$ counts per bin below $1$ keV, $\sim 10^3$ counts per bin in the Fe K band, and $\sim 500$ counts per bin at the highest probed energies.}. Because of the different count losses due to excision of the PSF in different observations, we were unable to recover the intrinsic average flux of the source, hence we do not cite normalisation values for the spectral fit referred to the stacked pn spectrum. As a sanity check, we also fit simultaneously the eight spectra of each individual observation; in this case, we can recover the intrinsic flux of the source by correcting the count rate for the PSF count loss. 
\begin{figure}[h!]
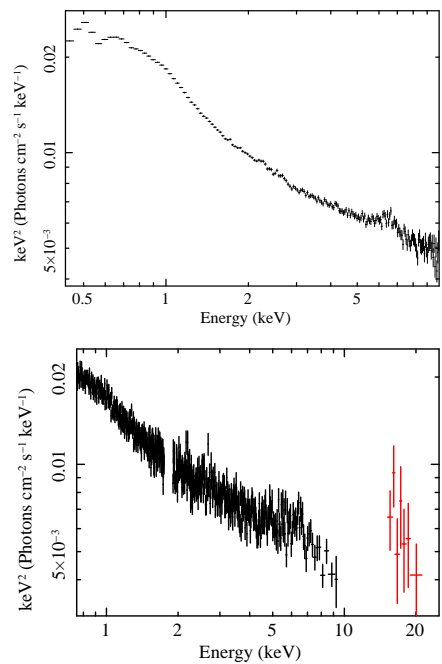

\centering
   \includegraphics[angle=-90,width=6cm]{figure4a.eps} 
   \includegraphics[angle=-90,width=6cm]{figure4b.ps} 
    \caption{ Observed frame $EF_E$ XMM-Newton EPIC pn (top) and Suzaku XIS+PIN (bottom) spectra  unfolded against a power law with $\Gamma=2$. \label{bbandEEUF}
    }
\end{figure}

\begin{figure}[h!]
   \includegraphics[angle=-90,width=7cm]{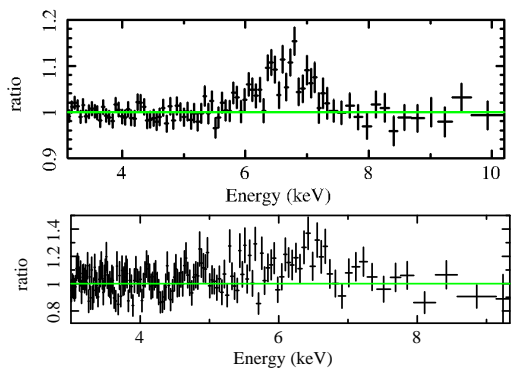} 
   \includegraphics[angle=-90,width=7cm]{figure5b.ps} 
    \caption{\label{FeK} Rest frame spectral residuals for a fit to a power law in the 3-6 and 7-10 keV band for the pn (top) and XIS (bottom) data.
    }
\end{figure}

The broad-band pn and XIS+PIN spectra are shown in Fig.~\ref{bbandEEUF} unfolded against a power law with $\Gamma=2$.
The spectra are extremely steep and show some structure in the Fe K band.
We first fitted the $3-10$~keV spectra to a simple power-law continuum emission. The fit statistics are $\chi^2/\nu=$484/512 for the XIS spectrum,  $\chi^2/\nu=190/113$ for the stacked pn spectrum, and $\chi^2/\nu=1084/905$ for the simultaneous fit to the eight individual pn datasets. 
Spectral residuals to this model are plotted in Fig.~\ref{FeK}, where the power law was fitted ignoring the Fe K region: even if a power law formally yields an excellent fit to the XIS data, significant residuals are evident.
Adding a Gaussian emission line significantly improved the fit for all the datasets. Spectral parameters for this model are reported in Table~\ref{T3-10}. The emission line centroid energy and width ($E_{line}\sim 6.6$ keV, $\sigma_{line}\sim 400$ eV) were found to remain constant within the errors in the different observations. The line equivalent width EW$_{line}$ was found to vary during the 2011 XMM-Newton observational campaign; it
was higher when the source flux was lower. We return to this point in Sect.~\ref{fluxsel} when we analyse the flux-selected pn spectra.

The single Gaussian line was replaced by the reflection model \texttt{xillver} \citep{2013ApJ...768..146G}, which simulates the reflected spectrum of a power-law incident on a plane-parallel optically thick slab. We used solar abundances and tied the slope of the incident continuum to that of the primary power law. Good fit statistics (respectively $\chi^2/\nu=$459/510, $112/111$, and $957/903$ for the XIS, the stacked pn, and the individual pn spectra) were obtained with a power law with $\Gamma\sim 2.4$ illuminating a highly ionised slab with $\log(\xi/$erg cm s$^{-1})=3.2-3.8$ (90\% c.l.). 
Allowing the normalisation of the reflected component to be a free parameter in the simultaneous fit to the eight individual pn spectra did not improve the fit. 

We then turned our attention to the full-band spectra, which were modelled including the three zones of photo-ionised gas and the three narrow emission lines detected in the RGS spectra (Sect.~\ref{xstar}). 
There is a prominent excess of emission both at lower energies (down to $E\sim 0.4$ keV) and at higher energies (up to 30 keV with Suzaku). 
The broad-band spectral shape of the high S/N stacked pn spectrum can be roughly modelled by a combination of \texttt{xillver} and either a power law plus a phenomenological blackbody ($\chi^2/\nu=607/181$), or \texttt{optxagn}, where the black hole spin was fixed to zero and the disk outer radius to $10^5r_g$ ($\chi^2/\nu=432/179$). 
However, a simple combination of \texttt{optxagn} plus an ionised reflector, as found for the fit to the $3-10$ keV data,  is unable to give a good fit to the broad-band data and leaves systematic residuals at $E\gtrsim 5$ keV.   
A further cold reflection component (modeled again with \texttt{xillver}) is able to well reproduce such high-energy residuals, but it strongly over-predicts the strength of soft emission lines at $E\lesssim 2$ keV. The addition of a layer of ionised gas with a column $\log (N_H/$cm$^{-2})\sim 23.5$ in front of the cold reflector gave a good fit to the broad-band data by excluding such strong emission lines (see Table \ref{broad}). 

\begin{figure}[h!]
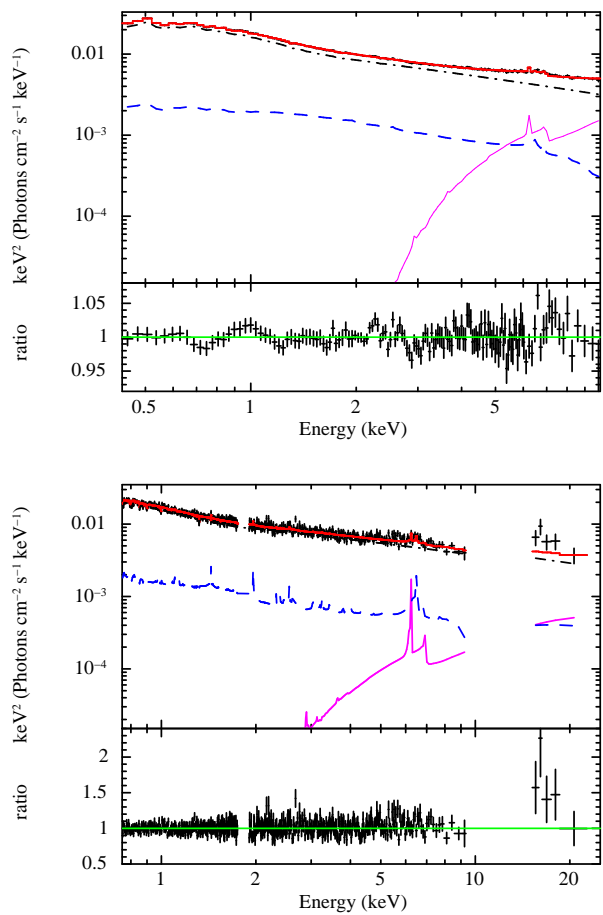

  \centering
    \includegraphics[angle=-90,width=0.45\textwidth]{figure6a.eps} 
    \includegraphics[angle=-90,width=0.45\textwidth]{figure6b.ps} 
      \caption{\label{figeuf} EPIC pn (top) and Suzaku XIS+PIN (bottom) spectra unfolded against the best-fit model described in Table~\ref{broad}, along with data-to-model ratios. The intrinsic continuum emission is modelled with \texttt{optxagn} (dotted-dashed black line); two reflectors are modelled with \texttt{xillver}, a highly ionised (dashed blue line) and a cold, absorbed one (solid magenta line). The total model is plotted in red. }
\end{figure}

The same model (\texttt{optxagn} + ionised reflection + cold and absorbed reflection) was able to well reproduce the XIS+PIN spectra without significant variations of the model parameters with respect to the stacked pn ones.
The observed variability for the eight individual pn spectra is driven by changes in the continuum amplitude. We found significant improvement of the fit statistics ($\Delta\chi^2/\Delta\nu=151/7$ and $\Delta\chi^2/\Delta\nu=76/7$, respectively) when allowing the photon index $\Gamma$ and the fraction of energy dissipated in the hot corona $f_{pl}$ to vary between the observations; these variations were very weak, however, with a maximum $\Delta\Gamma\sim0.1$ and $\Delta f_{pl}\sim 0.05$.  The best-fit model spectral parameters are reported in Table~\ref{broad}, while the Suzaku and stacked pn spectra are plotted in Fig. \ref{figeuf} unfolded against the best-fit model, together with data-to-model ratios.

\begin{table*}
\caption{Fit to the 3-10 keV band with a power law plus a Gaussian emission line\label{T3-10}}
\centering
\begin{tabular}{c c c c c c c c}
\hline\hline
 Dataset &  $\Gamma$ &  $E_{line}$/keV &  $\sigma_{line}$/eV & $I_{line}$ &  EW$_{line}$/eV &  $\chi^2/\nu$ &  $f_{3-10}$/10$^{-11}$ erg cm$^{-2}$ s$^{-1}$  \\
\hline
XIS & $2.50^{+0.14}_{-0.05}$ & $6.55^{+0.11}_{-0.10}$ & $231^{+234}_{-91}$ & $1.9^{+1.1}_{-0.7}$ & $138^{+80}_{-54}$ & 460/509  & $1.15^{+0.21}_{-0.08 }$\\
\multicolumn{8}{c}{}\\
pn stacked & $2.40^{+0.01}_{-0.02}$ & $6.58^{+0.09}_{-0.10}$ & $460^{+130}_{-100}$ & $1.6^{+0.4}_{-0.4}$ & $111^{+30}_{-26}$ & 96/110  & /\\
\multicolumn{8}{c}{}\\
201 & \multirow{8}{*}{$2.45^{+0.02}_{-0.02}$} & \multirow{8}{*}{$6.55^{+0.09}_{-0.10}$} & \multirow{8}{*}{$610^{+160}_{-120}$} & $<2.3$ & $<121 $&  \multirow{8}{*}{925/895} & $1.63^{+0.11}_{-0.10}$ \\
301 &   & & & $2.0^{+0.9}_{-0.7}$ & $165^{+89}_{-60}$ & & $1.06^{+0.10}_{-0.08 }$  \\
401 &  & & & $2.5^{+1.1}_{-0.9}$  &  $171^{+75 }_{-62}$ && $1.21^{+0.07}_{-0.06}$  \\
501 &  & &  & $2.5^{+1.1}_{-0.9}$ & $157^{+67}_{-59 }$ &&$1.37^{+0.08}_{-0.08}$   \\
601 &  & & & $2.4^{+1.0}_{-0.9}$  &  $166^{+68 }_{-57 }$ & & $1.24^{+0.07}_{-0.06}$   \\
701 &  && &  $2.9^{+0.9}_{-0.7}$ & $321^{+101 }_{-84 }$ && $0.77^{+0.06}_{-0.05}$  \\ 
801 &   && & $3.3^{+1.0}_{-0.9}$ & $229^{+70 }_{-60 }$ & & $1.24^{+0.07}_{-0.06}$   \\
901 &   & && $<1.9$ & $<108 $  &   & $1.45^{+0.10}_{-0.09}$\\
\multicolumn{8}{c}{}\\
pn low & \multirow{3}{*}{$2.45^{+0.02}_{-0.02}$} & \multirow{3}{*}{$6.57^{+0.09}_{-0.10}$} & \multirow{3}{*}{$\equiv 460$} &  $3.7^{+1.3}_{-0.9}$ & $245^{+33}_{-42}$ & \multirow{3}{*}{585/553} & $0.87^{+0.03}_{-0.02}$\\
pn med & & & & $3.1^{+1.2}_{-0.9}$ & $142^{+33}_{-34}$ & & $1.31^{+0.06}_{-0.04}$\\
pn high & & & & $<3.0$ & $< 81$ & & $1.95^{+0.07}_{-0.14}$\\
 \hline
 \end{tabular}
\tablefoot{ $I_{line}$ in $10^{-5}$ phot cm$^{-2}$ s$^{-1}$ - Xspec model: \texttt{powerlaw + gauss}.}
\end{table*}

\begin{table*}
\caption{Fit to the full band with optxagn, an ionised reflector, and an absorbed cold reflector\label{broad}}
\centering
\begin{tabular}{c c c c c c c c c c c c}
\hline\hline
 Dataset &  $\Gamma$ &  $L/L_{Edd}$  &  $r_{cor}/r_g$ &   $f_{pl}$ &  $kT$/eV &  $\tau$ & $\log\xi_{hot}$ &   $\log N_H$ &  $\log\xi_{abs}$ & $\log\xi_{cold}$ &  $\chi^2/\nu$  \\
\hline
XIS+PIN & $2.51\pm{0.02}$ & $1.13^{+0.04}_{-0.18}$ & $32\pm{2}$ & $0.73^{+0.04}_{-0.07}$ & $140^{+9}_{-5}$  & $61\pm{5}$ & $3.46^{+0.24}_{-0.07}$ &  $23.1^{+0.6}_{-0.3}$ & $<2.9$ & $<1.6$ & 1090/1109\\
\multicolumn{12}{c}{}\\
pn  & $2.59^{+0.02}_{-0.01}$ & $0.90^{+0.08}_{-0.13}$ & $>35$ & $0.75^{+0.02}_{-0.10}$ & $148^{+5}_{-6}$  & $41^{+15}_{-7}$ & $4.16^{+0.34}_{-0.43}$ &  $23.47^{+0.10}_{-0.19}$ & $<1.6$ & $0.86^{+0.06}_{-0.15}$ & 196/175\\
\multicolumn{12}{c}{}\\
201 & $2.63^{+0.02}_{-0.02 }$ & $1.46^{+0.2}_{-0.2 }$& \multirow{8}{*}{$58^{+32}_{-9}$} & $0.76^{+0.02 }_{-0.02 } $&  \multirow{8}{*}{$150^{+3}_{-2}$} &  \multirow{8}{*}{$38^{+29}_{-8}$} &  \multirow{8}{*}{$3.5^{+1.2}_{-1.1}$} &  \multirow{8}{*}{$23.65^{+0.07}_{-0.05}$} &  \multirow{8}{*}{$2.10^{+0.13}_{-0.16}$} &  \multirow{8}{*}{$<0.52$} &  \multirow{8}{*}{1491/1457}\\
301 &  $2.63^{+0.02 }_{-0.02 }$  & $1.02^{+0.02}_{-0.02}$ & &  $0.75^{+0.03}_{-0.03}$ & &   \\
401 & $2.57^{+0.02}_{-0.02}$  & $1.03^{+0.01}_{-0.01}$ & &  $0.76^{+0.03 }_{-0.03}$ &&  \\
501 & $2.59^{+0.02}_{-0.02}$  &$1.16^{+0.01}_{-0.01}$ & &  $0.75^{+0.02}_{-0.03 }$ &&   \\
601 & $2.58^{+0.02}_{-0.01}$  & $1.03^{+0.01}_{-0.02}$& &  $0.76^{+0.02 }_{-0.05 }$ & &    \\
701 & $2.59^{+0.03}_{-0.03}$  &$0.66^{+0.78}_{-0.02}$& &  $0.76^{+0.02 }_{-0.05 }$ &&    \\ 
801 &  $2.54^{+0.01}_{-0.02}$  & $0.99^{+0.03}_{-0.01}$ && $0.71^{+0.04 }_{-0.03 }$ & &    \\
901 & $2.63^{+0.02}_{-0.02}$   & $1.34^{+0.01}_{-0.02}$ &&$0.75^{+0.02}_{-0.03} $  &   &  \\
\multicolumn{12}{c}{}\\
pn low & \multirow{3}{*}{$2.66^{+0.03}_{-0.02}$} & $0.65^{+0.05}_{-0.06}$ & \multirow{3}{*}{$59^{+13}_{-25}$} & \multirow{3}{*}{$0.83^{+0.03}_{-0.02}$} & \multirow{3}{*}{$144^{+2}_{-3}$} & \multirow{3}{*}{$>51$} & \multirow{3}{*}{$3.91^{+0.61}_{-0.06}$} & \multirow{3}{*}{$22.74^{+0.09}_{-0.15}$} & \multirow{3}{*}{$1.97^{+0.19}_{-0.16}$} & \multirow{3}{*}{$<0.52$} & \multirow{3}{*}{1008/907} \\
pn med & & $1.03^{+0.03}_{-0.04}$ & & & & & & & & & \\
pn high & & $1.60^{+0.05}_{-0.07}$ & & & & & & & & &  \\
 \hline
 \end{tabular}
\tablefoot{Xspec model: \texttt{optxagnf + xillver$_{hot}$ + absorber*xillver$_{cold}$}. Units are cm$^{-2}$ for the column density $N_{H}$; erg cm s$^{-1}$ for the ionisation parameters $\xi_i$.}
\end{table*}

\begin{table*}
\caption{Flux-selected spectra}
 \centering
 \begin{tabular}{c c c c c c }
 \hline\hline
 Flux state &  OBSID  &  Exposure & Count-rate &  Stacked set  & Counts(4-10) \\
\hline
\multirow{4}{*}{Low} &  301 & 11.4 & $26.26\pm{0.05}$& \multirow{4}{*}{301+401+601+701}  & \multirow{4}{*}{25600} \\
          &  401 & 10.3 & $25.97\pm{0.05}$ & &   \\
          &  601 & 7.6 & $26.71\pm{0.06}$&  & \\
          &  701 & 24.8 & $20.41\pm{0.03}$&   & \\
                 \multicolumn{6}{c}{}\\
\multirow{5}{*}{Medium} &  301 & 9.5 &  $37.05\pm{0.06}$ & \multirow{3}{*}{301+401+601}& \multirow{3}{*}{18800}\\ 
&  401 & 9.4& $37.7\pm{0.06}$ & &   \\ 
&  601 & 8.2 & $37.21\pm{0.07}$ & &   \\ 
&  501 & 19.8 & $38.24\pm{0.04}$& 501& 9200 \\
&  801 & 23.2& $37.88\pm{0.03}$& 801& 11900 \\ 
                 \multicolumn{6}{c}{}\\
\multirow{2}{*}{High} &  201 & 12.4 & $57.16\pm{0.05}$ &  \multirow{2}{*}{201+901}& \multirow{2}{*}{10300}\\ 
&  901 & 9.7 & $61.41\pm{0.05}$ & &   \\ 
\hline
\end{tabular}
\tablefoot{Notes: flux state name, OBSID, exposure (including the live time), count-rate in the 0.4-10 keV band (corrected for the psf excision), stacked spectra, net counts in the 4-10 keV band.\label{fluxtable}}
\end{table*}

\paragraph{A blurred reflection scenario\label{BLUR}}

\begin{figure}[h!]
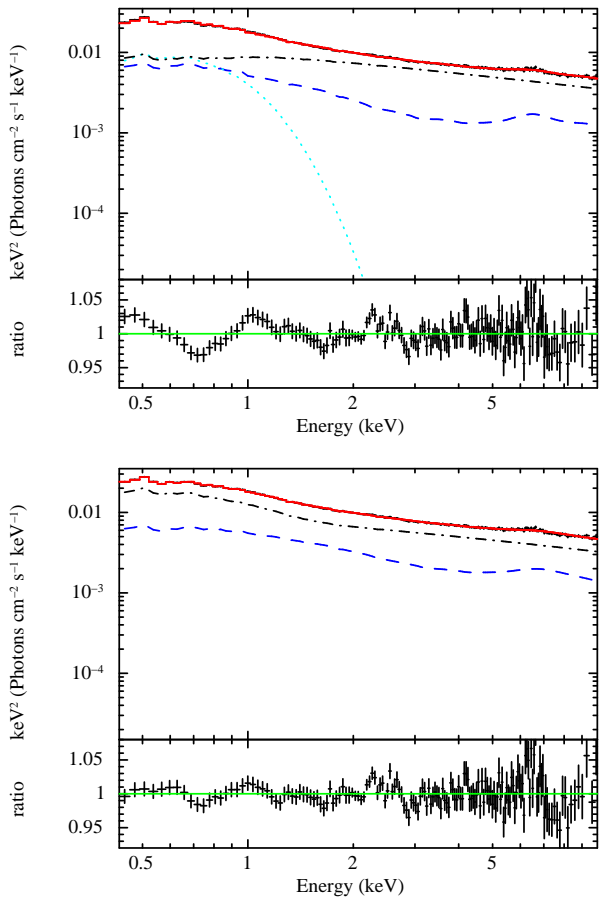

  \centering
    \includegraphics[angle=-90,width=0.45\textwidth]{figure7a.eps} 
    \includegraphics[angle=-90,width=0.45\textwidth]{figure7b.eps} 
      \caption{\label{kdbl} Stacked pn spectrum unfolded against two different representations of a blurred reflection scenario. Top panel: continuum modelled with \texttt{power law} (black dotted-dashed line) + \texttt{blackbody} (blue dotted line), and blurred reflection modelled with \texttt{kdblur*xillver} (blue dashed line); bottom panel: same as above, but modelling the continuum with \texttt{optxagn} (black dotted-dashed line).}
\end{figure}

We tested a blurred reflection scenario on the stacked pn spectrum. 
The intrinsic continuum was modelled with a power law plus a blackbody; a reflection component modelled with \texttt{xillver} was convolved with the relativistic smearing model \texttt{kdblur} \citep{1991ApJ...376...90L}. The emissivity index $q$ was fixed to 3 (a value typical of isotropic illumination of the reflecting slab) and the iron abundance was fixed to solar. A good fit ($\chi^2/\nu=197/179$) is obtained with a power law with $\Gamma=2.45\pm{0.01}$ reflecting onto a ionised ($\log(\xi/$erg cm s$^{-1})=2.98^{+0.02}_{-0.08}$) slab seen edge-on (inclination $i=89\pm 1$ deg), that extends close to a rapidly rotating black hole (with a slab inner radius $r_{in} < 1.85 r_g$ at 90\% c.l.); the unfolded spectrum is plotted in the top panel of Fig.~\ref{kdbl}. 

Replacing the \texttt{power law + blackbody} model with \texttt{optxagn} gave a good fit to the data as well ($\chi^2/\nu=167/177$) with the following parameters: an intrinsic X-ray continuum with  a slope $\Gamma=2.43\pm{0.01}$; a mass accretion rate $\dot{M}_{acc}=0.80^{+0.03}_{-0.02}\dot{M}_{Edd}$; a coronal radius $r_{cor}=37^{+3}_{-3} r_g$; a fraction $f_{pl}=0.58^{+0.03}_{-0.05}$ of accretion energy comptonized in the hot optically thin X-ray corona, the remaining $(1-f_{pl})$ Comptonized in a colder ($kT=152^{+1}_{-2}$ eV), optically thick ($\tau=32\pm{5}$) medium. The blurred reflection component has a high ionisation state ($\log(\xi/$erg cm s$^{-1})=2.91^{+0.11}_{-0.03}$) and extends down to $r_{in}<1.59 r_g$. 
The fraction of reflected or direct flux is $0.38^{+0.05}_{-0.12}$, $0.46^{+0.07}_{-0.14}$, and $0.43^{+0.06}_{-0.13}$ in the $0.4-1$, $1-4$, and $4-7.5$ keV band, respectively.
The corresponding unfolded spectrum is plotted in the bottom panel of Fig.~\ref{kdbl}.
A comparison between the top and bottom panels of Fig.~\ref{kdbl} shows the equivalent modelling of \texttt{optxagn} and a \texttt{power law + blackbody} where the peak temperature of the latter is $kT\sim 130$ eV. 

\begin{figure}[h!]
  \centering
    \includegraphics[angle=-90,width=0.45\textwidth]{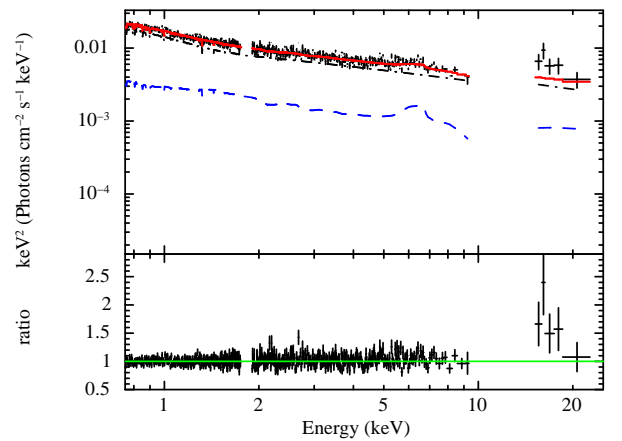} 
   \caption{\label{eufblur} Suzaku XIS+PIN spectra unfolded against a blurred reflection model, colour-coded as the bottom panel of Fig.~\ref{kdbl}. }
\end{figure}

The application of the blurred reflection scenario to the broad-band Suzaku data, with an emissivity index fixed to $q=3$ and iron abundances fixed to solar, gave a good fit ($\chi^2/\nu=1084/1109$) with spectral parameters compatible within the errors with those derived for the stacked pn spectrum, but for the inner radius of the reflecting slab, which is found to be much higher: $r_{in}=27^{+115}_{-20} r_g$. The XIS+PIN spectrum unfolded against this model is plotted in Fig.~\ref{eufblur}: despite the formally excellent fit, significant residuals in the PIN band are present, with the source flux being underestimated by the blurred reflection model. 

\paragraph{A partial covering scenario}
A partial covering scenario was tested by absorbing the intrinsic continuum (modelled with \texttt{optxagn}) with \texttt{zxipcf} \citep{2008MNRAS.385L.108R}. Both the stacked pn spectrum and the XIS+PIN spectra are well modelled by these two components ($\chi^2/\nu=198/178$ and $1095/1110$, respectively). The absorber column density, ionisation parameter, and covering fraction were $\log (N_H/$cm$^{-2})=23.62^{+0.08}_{-0.13}$, $\log(\xi/$erg cm s$^{-1})=1.0^{+0.3}_{-0.5}$, and $C_f=0.43^{+0.03}_{-0.05}$ for the pn, and $\log (N_H/$cm$^{-2})=23.75^{+0.20}_{-0.66}$, $\log\xi/$(erg cm s$^{-1})=1.9^{+0.3}_{-1.4}$, and $C_f=0.29^{+0.07}_{-0.06}$ for the XIS+PIN data. The absorbed component helps reproducing the spectral shape at high energies, basically replacing the absorbed reflector plotted as the magenta line in Fig.~\ref{figeuf}. 
The intrinsic continuum model parameters were similar for the two datasets and comparable with the parameters derived in the previous section and reported in Table~\ref{broad}, except for the mass accretion rate that in this scenario is estimated to be higher, $\dot{M}_{acc}/\dot{M}_{Edd}=1.83^{+0.11}_{-0.25}$ and $\dot{M}_{acc}/\dot{M}_{Edd}=1.51^{+0.05}_{-0.12}$ for the pn and XIS+PIN data. 

\subsubsection{Flux-selected spectra\label{fluxsel}}
\begin{figure}[h!]
  \centering
    \includegraphics[angle=-90,width=0.4\textwidth]{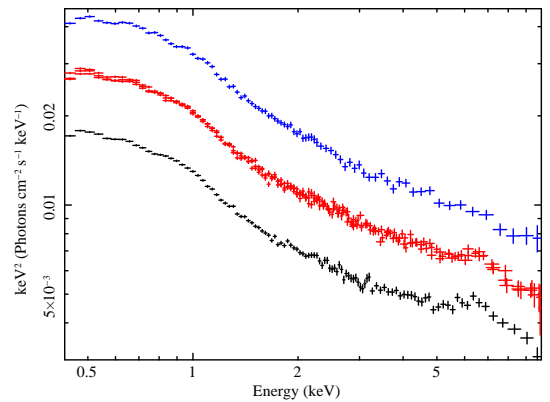} 
    \caption{\label{fsel}  $0.4-10$ keV flux-selected spectra unfolded against a power law with $\Gamma=2$: low (black), medium (red), and high (blue) flux states. }
\end{figure}
The EPIC-pn source spectra were stacked according to three different flux levels selected after inspecting Fig.~\ref{Flc}: a low flux state with a count rate $< 27$~counts~s$^{-1}$, a medium flux state with $30$~counts~s$^{-1} <$~count rate~$<40$~counts~s$^{-1}$, and a high flux state with a count rate $>45$~counts~s$^{-1}$.
Flux-selected spectra were then stacked according to the shape of the source extraction region and were fitted if they had $>5000$ net counts in the $4-10$~keV band. The relative contribution of the different observations to each flux-selected spectrum is reported in Table~\ref{fluxtable}, while the full-band unfolded spectra are plotted in Fig.~\ref{fsel}: despite flux changes of over a factor of two, the spectral shape does not vary dramatically. 

The $3-10$~keV band of the three flux-selected spectra was fitted simultaneously to a power-law model: the fit statistics was quite poor ($\chi^2/\nu=708/557$), and there were residuals in the Fe K band. As with the singular spectra, the addition of a Gaussian emission line (where the width $\sigma$ was fixed to 460~eV, i.e. the best-fit value found for the average stacked pn spectrum) significantly improved the fit ($\Delta\chi^2/\Delta\nu=123/4$). 
The line centroid energy is consistent with the one measured in the average spectra, but its equivalent width significantly decreases from lower to higher states (see Table~\ref{T3-10} for values). This suggests the presence of a constant (or varying slower than the continuum) component that emerges when the intrinsic continuum is in a low state. 

We applied the best-fit model found in Sect.~\ref{average} (\texttt{optxagn} + ionised \texttt{xillver} + absorbed cold \texttt{xillver}) to the flux-selected spectra, which yielded a good fit to the data ($\chi^2/\nu=1008/907$, Table~\ref{broad}) with the two reflection components remaining constant, and the observed variability being driven by variations in intrinsic continuum amplitude. 

\subsubsection{Time-resolved spectra\label{timeres_FD}}

\begin{figure}[h!]
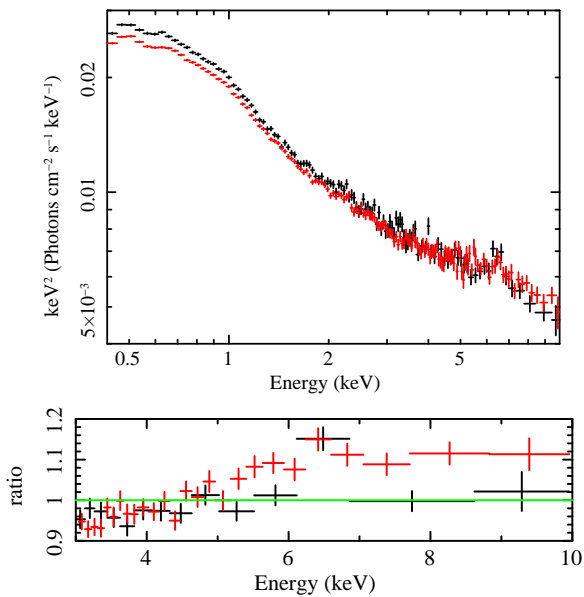

  \centering
    \includegraphics[angle=-90,width=0.4\textwidth]{figure10a.eps} 
    \includegraphics[angle=-90,width=0.45\textwidth]{figure10b.eps} 
   \caption{\label{tres} Top panel: soft flares (black) and their associated delayed signal (red) spectra unfolded against a power law with $\Gamma=2$. Bottom panel: data-to-model ratio to a power law fitted on the $2-10$~keV band.  }
\end{figure}

We aim at extracting the spectral signature of the 'hard delayed excess' that follows flares of emission in Ark 564, as identified by \citet{2012ApJ...760...73L}. To this end, we extracted the source spectra in time intervals centred on the position of the 'soft flares'\footnote{Soft flares are defined by \citet{2012ApJ...760...73L} as those only detected in the $0.4-1$ keV band.}, with a width of $\pm 384$~s. The associated delayed signal was extracted in a time interval $1088-2496$~s after each peak centre, that is, the time interval when the hard delayed excess appears to be the strongest \citep[see Fig. 6 of][]{2012ApJ...760...73L}.  All the 'soft flares' spectra were then stacked together, as were all the 'delayed' spectra. For the spectral analysis we used the same scaled background as we used to analyse the average stacked pn spectrum. Figure~\ref{tres} shows the unfolded broad-band spectra for the flaring (black) and their delayed (red) time intervals.
The variation in spectral shape is clear: while at energies $E\lesssim 2.5$~keV the spectral shape is just shifted down in intensity (as found for the flux-selected spectra), at $E\gtrsim 3$~keV there is an evident hardening of the delayed spectrum. This can be seen in the bottom panel of the same figure where high-energy residuals to a fit to a power law with $\Gamma\sim 2.4$ are plotted. 

We modelled the two spectra exploiting the constraints derived from the timing analysis  by  \citet{2012ApJ...760...73L} in terms of relative fractions of direct (flaring) and scattered (reprocessed) light. In particular, as the low-frequency lags are always positive with respect to the $0.4-1$ keV band, the fraction of scattered light must increase from the softest to the hardest band: we used this constraint to construct a model able to reproduce the flaring and delayed spectra.

\begin{figure}[h!]
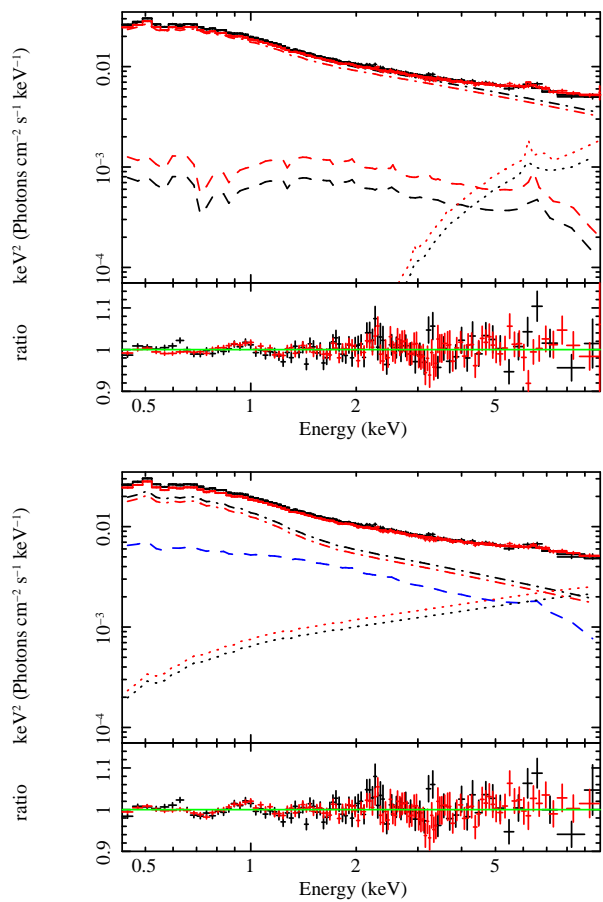

  \centering
    \includegraphics[angle=-90,width=0.45\textwidth]{figure11a.eps} 
    \includegraphics[angle=-90,width=0.45\textwidth]{figure11b.eps} 
   \caption{\label{tres2} Top panel: soft flares (black) and their associated delayed signal (black) spectra unfolded against a reflection-dominated model: \texttt{optxagn} (dotted-dashed lines), a heavily absorbed cold reflector (dotted lines), and a slightly absorbed hot reflector (dashed lines). Bottom panel: same as above, but unfolded against a Compton-upscattering-dominated model: \texttt{optxagn} (dotted-dashed lines), a hot reflector (blue dashed line), and a hot Compton-upscattering model (dotted lines). }
\end{figure}

First, we tested the best-fit model of Sect.~\ref{average} (i.e. \texttt{optxagn} plus an ionised reflector plus a cold, absorbed reflector, both modelled with \texttt{xillver}). 
Within this model, \texttt{optxagn} represents the direct light, while the two reflected components represent the scattered or
reprocessed light.  The normalisation of the continuum was allowed to vary between the two data sets and is $\sim 10\%$ higher in the flares than in the delayed spectrum. To match the constraints on the scattered or direct light fractions in the different energy bands, another layer of ionised gas was introduced in front of the ionised reflector. This model can reproduce the data very well ($\chi^2/\nu=350/355$) with continuum emission model parameters compatible with those measured for the average spectra (Sect.~\ref{average}, Table~\ref{broad}), but a higher flux in the two reflected components for the delayed signal: the 0.4-10 keV observed flux is $1.6^{+0.4}_{-0.8}\times 10^{-11}$ erg s$^{-1}$ cm$^{-2}$ and $1.6^{+0.1}_{-0.3}\times 10^{-12}$ erg s$^{-1}$ cm$^{-2}$  for the ionised and cold reflector in the delayed spectrum, respectively, as opposed to $1.3^{+0.6}_{-0.9}\times 10^{-11}$ erg s$^{-1}$ cm$^{-2}$  and $1.0^{+0.2}_{-0.4}\times 10^{-12}$ erg s$^{-1}$ cm$^{-2}$  in the flaring spectrum. The ionised reflector is absorbed by a layer of gas with $\log (N_H/$cm$^{-2})=21.6_{-0.1}^{+0.2}$ and $\log(\xi/$erg cm s$^{-1})=0.21^{+0.15}_{-0.14}$. The corresponding model and spectral residuals are reported in the top panel of Fig.~\ref{tres2}.

Alternatively, reprocessing may occur onto a hot medium such as to be able to Compton-up-scatter the input (soft flares) photons. We tested this scenario by replacing the cold absorbed reflector with the \texttt{comptt} model \citep{1994ApJ...434..570T}, where we assumed a disk-like geometry and a temperature for the seed photons equal to the temperature of the warm, optically thick medium responsible for the Compton-upscattering of disk photons inside $r_{cor}$ as derived from \texttt{optxagn}, that is, $kT^{optx}\sim 150$ eV. Within this model, \texttt{optxagn} again represents the direct light, while the scattered or reprocessed light is represented by \texttt{comptt}. 
 This model can also well reproduce the data ($\chi^2/\nu=380/360$) by allowing again the \texttt{optxagn} normalisation to be higher in the flare spectrum than in the spectrum with the delayed signal, and the Compton-upscattering component to emerge in the delayed spectrum, with a 0.4-10 keV observed flux of $5.6^{+0.2}_{-0.3}\times 10^{-12}$ erg cm$^{-2}$ s$^{-1}$ and $6.6^{+0.1}_{-0.2}\times 10^{-12}$ erg cm$^{-2}$ s$^{-1}$ for the flares and delayed spectrum, respectively. The temperature of the Compton-upscattering material is found to be $kT^{C}=4.3^{+0.8}_{-0.5}$ keV, its optical depth $\tau^C=6^{+3}_{-2}$. 
 The flares and delayed spectra unfolded against this model, as well as spectral residuals, are plotted in the bottom panel of Fig.~\ref{tres2}.

\paragraph{Detailed flare evolution\label{flarone}}
\begin{figure}[h!]
  \centering
    \includegraphics[width=0.5\textwidth]{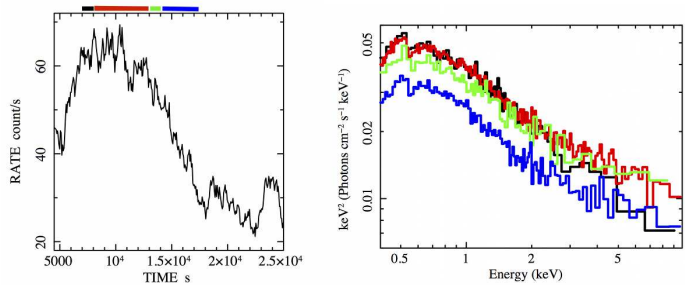} 
       \caption{ \label{901} Left panel: a segment of the 0.4-10 keV light curve of the pn observation 901 where a strong and sudden increase and decrease in flux is evident. Right panel: four 0.2-10 keV EPIC-pn spectral slices extracted at the beginning (during the first ks) of the high-flux level (black), during the remaining of the high-flux level (red), during the first ks of the decrease in flux (green), and during the remaining decreasing flux level (blue): the time intervals are marked in the top panel with the same colour-code. Error bars have been omitted for clarity. }
\end{figure}
We have shown that the rapid source variability is dominated by intrinsic continuum amplitude variations (Sect.~\ref{average}) and that a hard spectral component appears delayed with respect to the peaks of the emission (Sect.~\ref{timeres_FD}).
A plausible interpretation is that the continuum light reverberates on some reprocessing material located $\sim 1000 $ light-seconds away from the continuum source. If this is the case, we would expect to detect the reaction of this reprocessing gas following the longer-term source variability. 

To test this expectation, we selected a time interval during the XMM-Newton observation 901 where there is a strong long-term flaring emission with large amplitude, namely the 7000-17500 s segment. The total count rate undergoes amplitude oscillations of about a factor of two on a time scale of a few kiloseconds. In particular, there is a strong rise in flux just after the first 5 kiloseconds of the observation, when the flux increases by 50\% in about 1500 s. The flux remains on a high level for about  6 ks, then decreases by a factor $>2$ in about 5000 s (Fig.~\ref{901}, left panel). 
The corresponding spectra are plotted in the right panel of Fig.~\ref{901}, omitting the error bars for clarity: spectral variability in response to the flux changes is evident. We can parametrise the spectral hardness by means of the ratio $R$ of the observed count
rate in a hard ($4.5-7$ keV) and in a soft ($0.4-1$ keV) energy band. During the first kilosecond of the flare (black spectrum) the source is soft, with $R=0.012$; after one kilo second has elapsed, the hard band has had time to react to the increased soft flux (red spectrum) and is found to be in a medium hardness state, $R=0.017$. Then the flux starts to drop: during the first kilosecond of decreasing flux (green spectrum) the hardness is highest ($R=0.019$), given that the hard band still did not have time to adjust to the soft-band flux changes; then, after one  kilosecond has elapsed, the source is found again in a medium hardness state (blue spectrum, $R=0.017$).

\section{Results}\label{results}

We present the results of both time-averaged and time-resolved X-ray spectroscopy of eight XMM-Newton and one Suzaku observations of the NLS1 Ark 564. 

\begin{itemize}
\item The Ark 564 X-ray continuum appears extremely steep in the whole energy band analysed in this work ($0.4-30$ keV, see Fig.~\ref{bbandEEUF}). A fit to the RGS data in the range $0.4-1.9$ keV gives an intrinsic slope $\Gamma=2.5-2.8$ depending on the model adopted, \texttt{optxagn} or \texttt{(powerlaw + blackbody)} (Table~\ref{table2}). 
  Modelling the broad-band data with a more physical model (\texttt{optxagn + two reflectors}) reveals a hard spectral component emerging above $E\gtrsim 4$ keV; the intrinsic continuum slope of Ark 564 thus ranges between $\Gamma=2.5-2.7$ (Table~\ref{broad}).
\item High-resolution X-ray spectroscopy was performed in the soft band ($E<1.8$ keV) with the RGS instrument (Sect. \ref{RGS}), and revealed the presence of ionised gas along the line of sight in the form of moderately ionised C, N, O, and Fe (Fig.~\ref{Fs} and Table~\ref{Tgauss}). 
\item Modelling the absorption features with the 2D photoionisation code \texttt{Xstar} requires three layers of absorbing gas, with ionisation parameter $\log(\xi/$erg cm s$^{-1})\sim -0.8$, $1.3$, and $2.4$ and no velocity shift with respect to the systemic redshift of the source;  the total measured absorbing column ranges from $\sim 10^{21}$ cm$^{-2}$ (fully covering absorber) to $\sim 1.4\times 10^{21}$ cm$^{-2}$ (allowing the absorber to be partially covering the continuum source), see Fig.~\ref{Fxs} and Table~\ref{Txstar}. 
\item A weak feature (EW$\sim$ 150 eV, see Fig.~\ref{FeK}) is detected in emission in the Fe K band, and it can be modelled either with a single broad Gaussian ($\sigma\sim 300-600$ eV) or with three narrow Gaussian emission lines with energies fixed to 6.4, 6.7, and 6.97 keV. The gas responsible for creating such features has a  high ionisation state: replacing the Gaussians with a physically self-consistent reflection model such as \texttt{xillver} requires $\log(\xi/$erg cm s$^{-1})> 3.5$.
\item Overall, the broad band spectrum of Ark 564 can be well reproduced by the superposition over the intrinsic continuum (modelled with \texttt{optxagn}) of two reflected components, both modelled with \texttt{xillver}: one hot ($\log(\xi/$erg cm s$^{-1})\sim 3-4$), and one cold ($\log(\xi/$erg cm s$^{-1}) < 1.2$) and absorbed by a column density $\log (N_H/$cm$^{-2}) \sim 23-24$ (see Fig.~\ref{figeuf} and Table~\ref{broad}).
\item A blurred reflection scenario was also tested on the time-averaged spectra by convolving the reflected component with a relativistic convolution model (i.e. \texttt{kdblur}). Both the EPIC-pn and the Suzaku spectra can be reproduced statistically well in this scenario (Sect.~\ref{BLUR}, Fig.~\ref{kdbl}), albeit giving discrepant results in terms of the inner radius of the reflecting slab, which would be $r_{in} < 1.85 r_g$ in the case of the EPIC-pn, and $r_{in} = 27^{+115}_{-20} r_g$ in the case of the Suzaku data. This model underestimates the Suzaku PIN flux (see Fig.~\ref{eufblur}). 
\item A partial covering scenario is also able to well reproduce the EPIC-pn and Suzaku spectra, assuming that the intrinsic continuum is partially covered ($C_f \sim 0.4-0.3$) by a column of $\log (N_H/$cm$^{-2}) \sim 23.6-23.7$ of moderately ionised gas ($\log(\xi/$erg cm s$^{-1})\sim 1-2$), for the EPIC-pn and the Suzaku XIS+PIN data, respectively.
\item Given the highly variable X-ray flux of Ark 564 (see Fig.~\ref{Flc}), spectral analysis was performed on three flux-selected spectra (Sect.~\ref{fluxsel}). Despite variation in flux of a factor $>2$, the source spectral shape is does not vary dramatically (Fig.~\ref{fsel}). Applying the best-fit model for the average spectra (\texttt{optxagn} + \texttt{hot xillver} + \texttt{cold absorbed xillver}) to the flux-selected spectra yields a good fit to the data, with the observed variability found to be driven by intrinsic continuum amplitude variations.
\item Time-resolved spectral analysis was performed on the eight XMM-Newton datasets by stacking all the time intervals identified by \citet{2012ApJ...760...73L} as 'soft flares' as well as all the subsequent 'delayed' emission (Sect.~\ref{timeres_FD}): the 'delayed' excess of emission that follows the soft flares is spectroscopically identified as a spectral hardening at $E\gtrsim 4$ keV (Fig.~\ref{tres}).
\item The best-fit model for the time-averaged spectra was applied to the 'soft flares' and the 'delayed' spectra by using the additional constraints of the relative fractions of direct light (the 'soft flaring' signal) and scattered or reprocessed light (the 'delayed' signal) at different energies, as measured from the timing analysis by \citet{2012ApJ...760...73L} to 'tune' the different spectral components. 
\item In particular, from the notion that the fraction of scattered
or direct light must be decreasing from high to low energies, we added to this model another layer of ionised gas with $\log (N_H/$cm$^{-2})\sim 21.6$ and $\log(\xi/$erg cm s$^{-1})\sim 0.2$ in front of the ionised reflector. The model is thus \texttt{optxagn} + \texttt{hot absorbed xillver} + \texttt{cold absorbed xillver}. Within this model the observed spectral variability is driven by a decrease in continuum amplitude and an increase of the reflected flux  in the 'delayed' spectrum compared to the 'flaring' spectrum. Given the time scales involved, a plausible physical scenario is to assume that the soft photons emitted during the continuum flare interact with the atoms of a gas located $\sim 100 r_g$ from the continuum source and are scattered, which produces the hard 'delayed' emission.
 \item The gas responsible for scattering the soft flare emission could also be so hot as to be extremely ionised and thus be
able to Compton up-scatter the flaring soft photons. We tested this scenario by replacing the cold absorbed reflector with the Comptonization model \texttt{comptt} and by removing the layer of absorbing gas in front of the ionised reflector (given the spectral shape of   \texttt{comptt}, it is not necessary in this scenario). Again, we tuned the model following the constraints of the relative fractions of direct or scattered light in different energy bands. We found that this model is also able to reproduce the data with respect to the scattered or direct light fraction constraints in all energy bands,if the gas has a temperature $kT\sim 4$ keV and a moderate optical depth $\tau\sim 6$ (Fig.~ \ref{tres2}). 
 \item A qualitative analysis of the detailed evolution of a single strong flare was presented. We showed that the source spectral shape hardens $\sim 1$ ks after a rise in flux, and that conversely, after a drop in flux it remains hard for $\sim 1$ ks before returning to its 'usual' spectral shape (Fig.~\ref{901}).
This is hard to reconcile with relativistically blurred reflection-dominated scenarios because we would expect the soft and hard band to 'communicate' on much shorter time scales (corresponding to a few light travel times instead of to $\sim$ a hundred), and strongly points, again, to reprocessing of the continuum photons by ionised gas located at 10s-100s$\,r_g$ from the central SMBH.
\end{itemize}

\section{Discussion and conclusions\label{concl}}

Ark 564 has a very steep X-ray continuum: a simple power-law model fit to the Suzaku and EPIC-pn data above $3~$keV yields  a slope of $\Gamma\sim 2.4$ (Table~\ref{T3-10}). However, the intrinsic continuum is even steeper, being observationally diluted by a reflected or reprocessed component that emerges in the hard band above $E\gtrsim 4$ keV both in the time-averaged and in the time-resolved spectra. In a standard two-phase model for the X-ray emission of AGN  \citep[a cold optically thick accretion disk + a hot optically thin corona,][]{1991ApJ...380L..51H, 1993ApJ...413..507H}, such a steep intrinsic continuum would suggest a high optical depth ($0.5 \lesssim \tau <1$) and/or a low temperature ($\Theta\lesssim 50$~keV) for the hot X-ray corona \citep{1997ApJ...476..620H}.

To first order, the broad band spectra of Ark 564 are well modelled either by  \texttt{(power law + blackbody)} or by \texttt{optxagn} and do not vary dramatically in shape despite flux variations of the order of a factor of $\sim 2$ (Fig.~\ref{fsel}).
There are clear signs of reprocessing gas in the inner regions of Ark 564, both in the soft and in the hard X-ray band.

In the soft band, a low column density of moderately ionised gas was found in the form of three different ionisation zones; the lowest ionisation state component of this absorbing gas might be the same as was observed in the UV band and studied by \citet{2002ApJ...566..187C}, who measured a modest velocity shift $\upsilon_{UV}\sim 200$ km s$^{-1}$ and a column density $N_{H,UV}\sim 1.5\times 10^{21}$ cm$^{-2}$. Three zones of ionised gas, with a comparable total column density as measured in this work, as well as a modest (an upper limit of $\upsilon_{out} < 110$ km s$^{-1}$) outflowing velocity for the warm absorber was also found by \citet{2008A&A...490..103S} in the analysis of five stacked XMM-Newton observations of Ark 564 performed between 2000 and 2001. A Chandra/HETG observation performed in 2000 also revealed multiple phases of ionised gas with a modest column density and zero net velocity shift \citep{2004ApJ...603..456M}; new observations performed with the same instrument in 2008 resolved a velocity of about 100 km s$^{-1}$ for the multiphase absorber \citep{2013ApJ...768..141G}.
Using the same dataset, \citet{2013ApJ...772...66G} claimed the discovery of a mildly relativistic ($\upsilon_{out}\sim 0.1 c$) outflowing absorber by identifying three absorption features due to OVI and OVII. We note that the RGS data show two  non-identified absorption features (Fig.~\ref{Fs}), which, if identified with NVII and OVII K$\alpha$ transitions, would suggest a similar outflow velocity $\upsilon_{out}\sim 0.1 c$. However, the identification of a wind phase based solely on two absorption troughs where most reasonable photo-ionisation models would predict a wealth of absorption lines due to many species is challenging, and we do not discuss these features further.

In the hard X-ray band, a weak feature (EW$\sim$ 150 eV) is detected in emission in the Fe K band, and it can be modelled either with a single broad Gaussian ($\sigma\sim 300-600$ eV) or with three narrow Gaussian emission lines with energies fixed to 6.4, 6.7, and 6.97 keV. In any case, ionised iron in the inner regions of Ark 564 is required by the data.
Given the width of the single Gaussian, simple Keplerian motion of a distribution of clouds around the SMBH would suggest a distance of the order of 100-1000s $r_g$. The same range of distance would be suggested by the lack of reaction of the iron emission complex to the large amplitude variations of the ionising flux (Sect. \ref{fluxsel}).  

However, to reproduce the spectral shape at the highest energies probed in this study, a hard-dominated component is required. This can be modelled either with a second (colder) reflector, a partially covered component, or with a relativistically blurred reflector. In the cold reflection scenario,
 a column $N_H>10^{23}$ cm$^{-2}$ must absorb the reflected emission
to not over-predict the strength of soft X-ray emission lines, which results in a quite contrived geometry. In the partial covering scenario, during the four years elapsed between the Suzaku and XMM-Newton observations the absorber covering factor would have significantly decreased while its ionisation state would have increased; in addition, the mass accretion rate in this scenario would be super-Eddington, which is unlikely. In the blurred reflection scenario, the XMM-Newton data require an accretion disk extending down to $<2r_g$ of the black hole, which is inconsistent with  the Suzaku data, which require $r_{in}\sim 30 r_g$. Noting also how the inclination angle of $\sim 90$ degrees would be hard to reconcile with the fact that Ark 564 is seen almost face-on \citep{2008A&A...490..103S}, and the severe underestimation of the PIN flux in this scenario (Fig.~\ref{eufblur}), we do not consider this to be physically plausible. \citet{2013MNRAS.434.1129K} reported the discovery of high-frequency Fe K lags of about 100 s in the first half of the 2011 XMM-Newton observational campaign on Ark 564 and ascribed them to relativistic reflection close to the central SMBH. It is unclear, however, how to reconcile this scenario with the lack of any lag during the second half of the observational campaign. Relativistic effects might be at work in the inner regions of Ark 564, but probably not that extreme: given the black hole mass estimate for Ark 564, a time lag of 100 seconds corresponds to about ten gravitational radii. \citet{2014MNRAS.439.3931E} inferred in the framework of a relativistic reflection lamp-post model a low spin of about 0.05 for the central SMBH of Ark 564, which again suggests that relativistic effects, if present, are not dominant in the very inner regions of Ark 564. 

Alternatively, reprocessing of the soft photon flares could occur in a medium so hot as to be able to Compton-upscatter them, giving rise to the hard excess component. In this case, we would have three different regions where Compton-upscattering occurs: a fraction (about 75\%) of the UV seed photons from the accretion disk within $\sim 50 r_g$ from the central supermassive black hole would be Comptonized in the very hot and optically thin X-ray corona ($\tau <<1, kT\sim 100$ keV), giving rise to the primary X-ray power-law emission; the remaining $\sim 25\%$ of the UV seed photons would be Comptonized in a warm and optically thick medium ($\tau\sim 40, kT\sim 150$ eV) that contributes to the emission below $\sim 2$ keV; and all of this Comptonized emission would be then further Compton-upscattered in a hot and optically thick medium ($\tau \sim 6, kT \sim 4$ keV) located 10s to 100s of gravitational radii from the central supermassive black hole, following flares of emission.
It would then follow that Comptonization plays a fundamental role in explaining the physics of the inner regions of AGN: 
in recent years, evidence is growing for it to explain the broad-band spectrum of several AGN, see, for instance, Mrk 509 \citep{2011A&A...534A..39M,2013A&A...549A..73P}, PG 1244+026 \citep{2013MNRAS.436.3173J}, 1H 0419-577 \citep{2014A&A...563A..95D}, Ark 120 \citep{2014MNRAS.439.3016M}, and NGC 5548 \citep{2015arXiv150101188M}, and references therein. 
Future high-quality spectral and timing observations on extended energy ranges with satellites such as NuSTAR \citep{2013ApJ...770..103H} will be able to distinguish between the different physical scenarios and test the hypothesis that Comptonization is the main mechanism at work.

\begin{acknowledgements}
      We acknowledge the referee for their useful comments that helped improving the manuscript.
      MG would like to thank Hiroki Akamatsu for useful discussions, and the NWO and the ESA Research Fellowship in Space Science program for financial support. 
      Based on observations obtained with XMM-Newton, an ESA science mission with instruments and contributions directly funded by ESA Member States and NASA.
      This research has made use of the NASA/IPAC Extragalactic Database (NED), which is operated by the Jet Propulsion Laboratory, California Institute of Technology, under contract with the National Aeronautics and Space Administration, and of the NASA Astrophysics Data System Bibliographic Services.
            
\end{acknowledgements}

\bibliographystyle{aa} 
\bibliography{a564.bib}
 
\end{document}